\documentclass[aps,twocolumn, superscriptaddress, preprintnumbers]{revtex4}
\usepackage{graphicx}
\newcommand*\diff{\mathop{}\!\mathrm{d}}
\newcommand\subscr[1]{_{\textrm{\scriptsize #1}}}

\newcommand\wa{0.95}
\newcommand\wb{0.9}

\begin{document}
\preprint{submitted for publication in \textit{J. Appl. Cryst.}}

\title{Exact resolution function for double-disk chopper\\ neutron time-of-flight spectrometers\,:
Application to reflectivity}

\author{Didier Lairez}
\affiliation{Laboratoire L\'eon Brillouin, CEA-CNRS-Universit\'e Paris-Saclay, 91191 Gif-sur-Yvette, France}
\affiliation{Laboratoire des solides irradi\'es, CEA-\'Ecole polytechnique-CNRS-Institut Polytechnique de Paris, 91128 Palaiseau, France}
\email{lairez@cea.fr}

\author{Alexis Chennevi\`ere}
\affiliation{Laboratoire L\'eon Brillouin, CEA-CNRS-Universit\'e Paris-Saclay, 91191 Gif-sur-Yvette, France}

\author{Fr\'ed\'eric Ott}
\affiliation{Laboratoire L\'eon Brillouin, CEA-CNRS-Universit\'e Paris-Saclay, 91191 Gif-sur-Yvette, France}

\date{\today}

\begin{abstract}
The exact resolution function in transfer vector for the reflectometer HERM\`ES at Laboratoire L\'eon Brillouin is calculated as an example  of neutron time-of-flight spectrometer with double-disk chopper. Calculation accounts for wavelength distribution of the incident beam, tilt of the chopper axis, collimation and gravity, without approximation of gaussian distributions or independence of these different contributions. Numerical implementation is provided that matches the sections of the paper. We show that data fitting using this exact resolution function allows us to reach much better results than its usual approximation by a gaussian profile.
\end{abstract}

\maketitle

\section{Introduction}

It is quite common for physical measurements to attempt to reach the limit of a given technique. In many cases, this amounts to measure a quantity with an accuracy better than the resolution of the apparatus. Said like that, it seems impossible. For instance, how to discriminate the position of two peaks which are closer from each other than their width\,? It is possible if we expect a given shape for each peak, do the convolution of expectation with the resolution function of the apparatus and compare the result with the measurement (this is commonly called ``data fitting''). Another example\,: neutron specular reflectivity allows us to get structural informations on thin layers at an interface \textit{i.e.} layers thicknesses and densities (for an introduction to reflectivity see for instance \cite{Cousin_2018}). 
The latters are determined relative to the scattering length density difference $\Delta\rho$ between the two infinite media separated by the interface and consequently from the value of the critical transfer vector $q_c$ below which total reflection occurs; \textit{e.g.} for the air/silicon interface $ q_c=0.0102\textrm{ \AA}^ {-1}$ yields to $\Delta\rho= q_c^2/16\pi = 2.07\times 10^{10} \textrm{cm}^{-2}$, which is the correct result. A shift of 3\% for $q_c=1.03\times 0.0102\textrm{ \AA}^ {-1}$ gives $\Delta\rho= 2.20\times 10^{10} \textrm{cm}^{-2}$, which is not acceptable for many users of reflectometers. But 3\% is the order of magnitude of the resolution. This barrier can be bypass if we know that the reflectivity curve should obey to a given function. To reach these limits, the calculation of the resolution function has to be as accurate as possible. In particular the approximation that all random variables that contribute to the  resolution have a gaussian probability density is likely not satisfactory\,\cite{nelson_towards_2013} especially in case the resolution is relaxed to gain flux. This approximation, which was legitimate when the means of calculation were insufficient, is no longer justified.

Neutrons reflectivity measurements take a ``picture'' of a sample in the reciprocal space for which the conjugate variable of distance is the transfer vector $q$ that is practically proportional to the ratio of the incident angle $\theta$ to the wavelength $\lambda$. Assuming that $\theta$ and $\lambda$ are two independent variables, the relative resolution is such as $(\sigma_q/q)^2\simeq(\sigma_\theta/\theta)^2+(\sigma_\lambda/\lambda)^2$, where $\sigma$ holds for standard deviations. On the other hand, if we do not account for the transfer function of the sample, the signal is proportional to the incident neutron-flux \textit{i.e.} to the product $\sigma_\theta \sigma_\lambda$. Thus for a given flux, the minimum for $\sigma_q/q$ is obtained for $\sigma_\theta/\theta=\sigma_\lambda/\lambda$. 
Time-of-flight techniques do the work at constant $\theta$ (thus constant $\sigma_\theta/\theta$) as a function of $\lambda$. If we require a constant flux, since $\sigma_\theta/\theta$ is constant, $\sigma_\lambda/\lambda$ should also be constant. This is achieved with double-disk choppers\,\cite{van_well_double-disk_1992} and this argument is the major reason why double-disk chopper are widely used. The counterpart is a broad resolution $\sigma_\lambda$ at large wavelength, precisely in the region where a good accuracy is often needed (edge of the total reflection plateau). Hence the interest of making an exact calculation of the resolution function.

In this paper we present the calculation of the overall and exact resolution function for the reflectometer HERM\`ES at Laboratoire L\'eon Brillouin which design is wide spread over neutron sources. In spite its current neutron source is continuous (Orph\'ee reactor), this spectrometer is based on the time-of-flight principle that is to become generalized with the increase of pulsed neutron sources, it is equipped with a double-disk chopper that is now a standard. The presented formalism to get the exact resolution function has thus a broad scope and could be easily transfered to other reflectometers or to other techniques such as small angle scattering or diffraction.

For the most part, the different contributions to the final resolution have already been mentioned separately in the literature\,: wavelength distribution of the incident beam, tilt of the chopper axis, beam size, collimation and gravity \cite{van_well_resolution_2005, nelson_towards_2013, gutfreund_towards_2018}. However considering all these different terms for the exact and ``all-in-one'' resolution function requires special attention, because the different contributions are neither independent, nor gaussian, they cannot be simply added and are quite intertwined.
This is precisely the motivation of our paper.
Here, the different random variables that contribute to the final resolution are presented in a uniform and comprehensive manner. But most importantly is that their variances are not just added in virtue of the central-limit theorem, as it is usually done, but their exact distribution functions are considered and appropriately convolved. We write analytically the whole resolution function that accounts for all terms without any approximation of gaussian distributions or statistical independence. In addition, a python code that implements numerically these calculations is provided (\url{https://bitbucket.org/LLBhermes/pytof/}). 
Finally, we show that data fitting using this exact resolution function allows us to reach much more accurate results than its usual approximation by a gaussian profile.

\section{Brief description of the reflectometer}

\begin{figure}
\centering
\includegraphics[width=\linewidth]{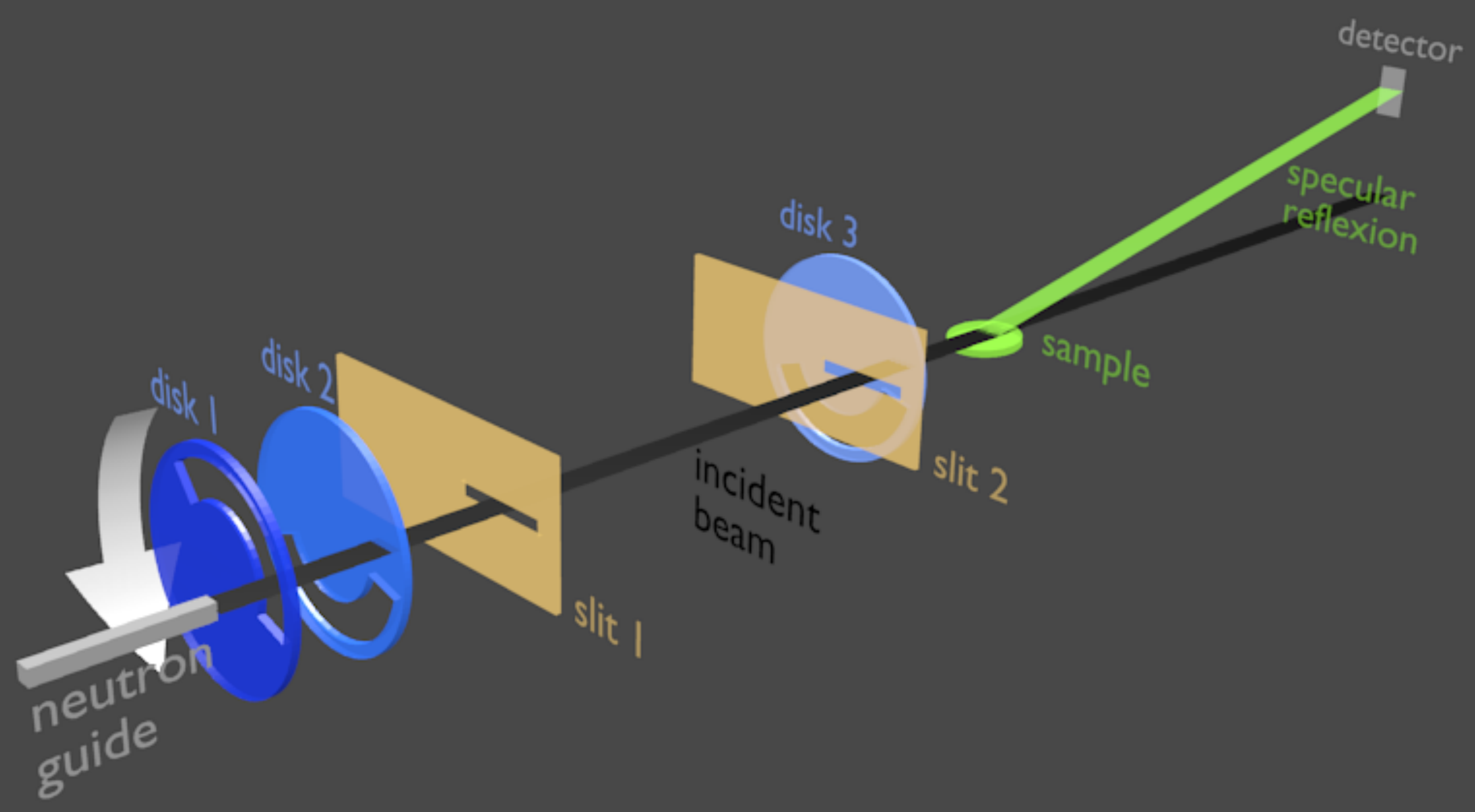} 
\caption{Schematic diagram of HERM\`ES reflectometer.}
\label{spectro}
\end{figure}

In Fig.\ref{spectro}, a schematic diagram of the reflectometer HERM\`ES is shown. Neutron-pulses are produced by a three-disk chopper from Airbus company. The three disks (numbered 1, 2 and 3 with respect to neutrons direction) have same radius $r=300$\,mm  and same fixed angular aperture equal to $165^\circ$ allowing neutrons to pass. In standard configurations these disks rotate in the same direction at a pulsation $\omega$ around the same axis. Disk~2 and 3 are in a fixed position 2\,m from each other, whereas disk~1 can be placed at three different distances from disk~2 (0.1, 0.35 and 1\,m, respectively). Essentially, disk~1 and 2 control the wavelength resolution whereas disk~3 in a standard configuration is mainly devoted to avoid the time-overlap of the slowest neutrons of a given pulse with the fastest of the next one.

The collimator is basically made of two horizontal slits of half-width $r_1$ and $r_2$ (numbered 1 and 2 with respect to neutrons direction) located between disk 2 and 3, at the same height and spaced by $d_c=1.8$\,m. The width of the slits are tuned so that the angular resolution remains consistent with the wavelength resolution resulting from the disk chopper parameters. In the following, we will consider three different typical configurations, such as $\sigma_\theta/\theta\simeq\sigma_\lambda/\lambda$, which correspond to high (HR), medium (MR) and low resolution (LR), respectively (see Table \ref{tab1}).

\begin{table}
\caption{Typical configurations corresponding to high (HR), medium (MR) and low resolution (LR). $x_1$, $x_2$, $x_3$ and $x_d$ are the positions of disk 1, 2, 3 and of the detector, respectively. $r_1$ and $r_2$ are the half-widths of the first and second slit of collimator at fixed distance $d_c=1800$\,mm apart. All lengths are in mm. Except when specified, the number of time-of-flight channels is equal to 256 covering 360$^\circ$ of chopper revolution. Also, the phases for disk 1 closure ($\varphi_{c,1}$), disk 2 and disk 3 openings  ($\varphi_{o,2}$, $\varphi_{o,3}$) are equal.}
\begin{center}
\begin{tabular}{r|c|c|c|}
&HR & MR & LR\\
$x_d-x_3$ & \multicolumn{3}{c|}{2375}\\
$x_3-x_2$ &  \multicolumn{3}{c|}{2000}\\
$x_2-x_1$ & 100 & 350 & 1000\\
$r_1$ & 0.5 & 1 & 2.3\\
$r_2$ & 0.5 & 0.5 & 0.5\\
\end{tabular}
\end{center}
\label{tab1}
\end{table}

Specular reflection at the desired angle $\theta_0$ is obtained by rotation of the sample. In case a non-horizontal beam is needed (\textit{e.g.} for the study of horizontal liquid surfaces), two plane mirrors are placed in the collimator to deviate the beam. These mirrors have no incidence on the resolution and will be ignored in the following. Specular reflection is measured in a vertical plane at angle $2\theta_0$ with a single detector whose area is wide enough to cross the trajectory of all reflected neutrons.

\section{Wavelength resolution}

The wavelength resolution results basically from the incident beam distribution and from the transfer function of the chopper. The latter is mainly controlled by the phases of the first two-disks, but in a general way the third also should be accounted for at long wavelength. In this section we examine these different points and present the way to calibrate the phases in question.

\subsection{Basics of double-disk chopper}

\begin{figure}
\centering
\includegraphics[width=\wb\linewidth]{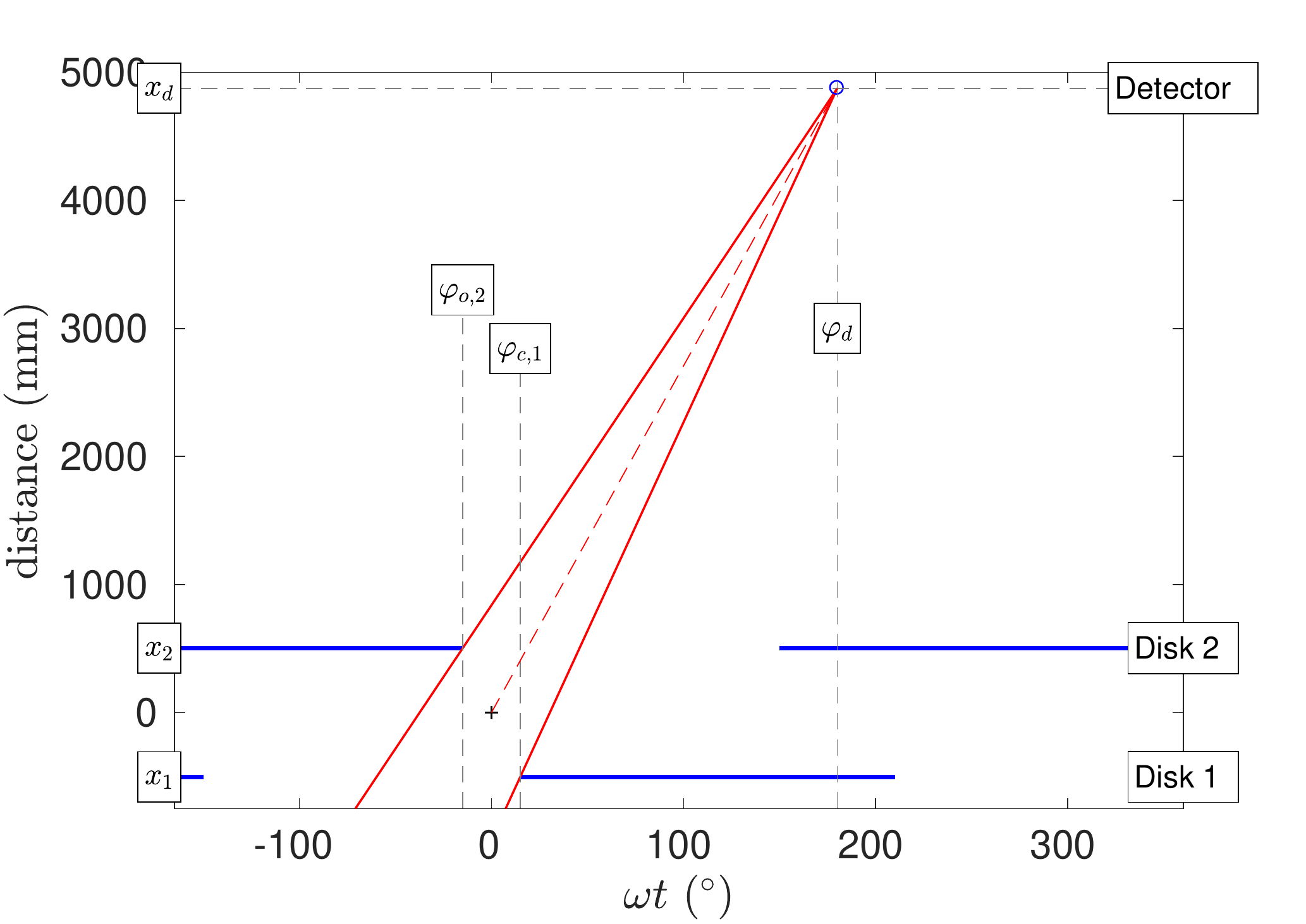} 
\caption{Double chopper: flight distance vs. phase $\omega t$. In blue the closed sector of disks. Neutrons that reach the detector at phase $\omega t_d$ have a kinematics that lies between the two red lines which slopes are given by Eq.\ref{velocity}.}
\label{chopper1}
\end{figure}

We first consider a chopper made only of the first two disks. Let us denote $x_k$ the positions (as in Table \ref{tab1}), $\varphi_{o,k}=\omega t_{o,k}$ the phase for disk opening and $\varphi_{c,k}=\omega t_{c,k}$ the phase for disk closure. By convention, we denote the actual phase as $\omega t-\varphi$, so that $\varphi>0$ states for a delay. The measurement consists in recording the number of neutron arrivals at detector position $x_d$ and time $t_d$ over a timebase that is periodically restarted (triggered) at each revolution of the chopper. The cumulated record, obtained at time $t_d$ is referred to as a ``time-of-flight channel'' (or tof-channel) in the following.
Fig.\ref{chopper1} shows the corresponding time-of-flight diagram using $\omega t$ as abscissa. 
From the de~Broglie's equation, in this diagram the kinematics curve of a neutron of wavelength $\lambda$ and velocity $v$ is a straight-line with the reciprocal slope
\begin{equation}
\frac{\omega}{v}=\frac{\omega}{h_m} \times\lambda
\end{equation}
where $h_m$ is the ratio of the Planck's constant to the mass of neutron\,: $h_m\simeq 3956\textrm{\,\AA}$\,m/s.

Simple geometric considerations show that neutrons with time of arrival $t_d$ have a velocity between $v_1$ and $v_2$, symbolized by the two red lines in Fig.\ref{chopper1}, such as
\begin{equation}\label{velocity}
\frac{1}{v_1}=\max\left\{{0, \frac{t_d  -t_{c,1}}{L_1}}\right\}
\quad\textrm{and}\quad
\frac{1}{v_2}=\max\left\{{0, \frac{t_d  -t_{o,2}}{L_2}}\right\}
\end{equation}
with $L_1=x_d-x_1$ and $L_2=x_d-x_2$. To these boundaries correspond the wavelengths $\lambda_1= h_m/v_1$ and $\lambda_2= h_m/v_2$. Let us define

\begin{equation}\label{red1}
\begin{array}{c}
L=(L_1+L_2)/2 \\
t_0=(t_{o,2} + t_{c,1})/{2}\\
l=(x_2-x_1)/2 \\
\epsilon=(t_{o,2} - t_{c,1})/{2}
\end{array}
\end{equation}
Note that $\epsilon$ can be either positive or negative ($\epsilon<0$ in Fig.\ref{chopper1}).
Then the second terms under braces in Eq.\ref{velocity} rewrite
\begin{equation}\label{velocity2}
\frac{1}{v_1}=\frac{(t_d-t_0)  +\epsilon }{L_1}
\quad\textrm{and}\quad
\frac{1}{v_2}=\frac{(t_d-t_0) -\epsilon}{L_2}
\end{equation}
The half range of transmitted wavelength $\Delta\lambda$ and the median wavelength $\lambda_0$ are\,:
\begin{equation}\label{median}
\Delta\lambda=({\lambda_2-\lambda_1})/{2}
\quad\textrm{and}\quad
\lambda_0=({\lambda_2+\lambda_1})/{2} 
\end{equation}
In the following, $\lambda_0$ will be reffered to as the nominal wavelength of the time-of-flight channel at time $t_d$. One gets\,:
\begin{equation}\label{resolChopper0}
\Delta\lambda=h_m\frac{l(t_d-t_0)-L\epsilon}{L_1L_2}
\quad\textrm{and}\quad
\lambda_0=h_m\frac{L(t_d-t_0)-l\epsilon}{L_1L_2}
\end{equation}
In the case $\epsilon=0$, one obtains\,:
\begin{equation}\label{epsilon0}
{\Delta\lambda}/{\lambda_0}={l}/{L}
\end{equation}
so that the relative resolution of the chopper is constant whatever the time-of-flight channel. This is the main reason for using double-disk chopper as it allows us to optimize the resolution with respect to neutron-flux\,\cite{van_well_double-disk_1992}. From Table\,\ref{tab1}, the three standard configurations correspond to ${\Delta\lambda}/{\lambda_0}=1$, 4 and 10\%, respectively. For comparison with other terms that play a role in the resolution, it is valuable to introduce a "pulse half-width" $\omega\tau_p=\Delta\lambda \times \omega L/h_m$.
For $\omega=10^4$\, $^\circ\textrm{s}^{-1}$, $\epsilon=0$, $L\simeq4$\,m and  $2\textrm{\AA}<\lambda<20\textrm{\AA}$,
one has $0.2^\circ<\omega\tau_p<2^\circ$ at high resolution and $2^\circ<\omega\tau_p<20^\circ$  at low resolution.

\begin{table}[!htbp]
\caption{Table of symbols}
\begin{center}
\begin{tabular}{r|l}
$h_m$ & ratio of the Planck's constant to the mass of neutron\\
$g$&gravitational acceleration\\ \\
$v$& neutron velocity\\
$\lambda$& wavelength\\
$\omega$& chopper pulsation\\
$t$& time\\
$\varphi$&$=\omega t$\\
$\delta$&tilt angle of disk1 - disk2 axis\\
$\eta$& tilt angle of disk1 - disk3 axis\\
$t_{c,k}$, $t_{o,k}$& closure and opennig time for disk $k\in\{1,2,3\}$\\
$t_d$& detection time (center of tof-channel)\\
$x_k$& position of disk $k\in\{1,2,3\}$\\
$x_d$, $x_c$, $x_s$& position of detector, middle of collimator, sample\\
$x\subscr{in}$& position of collimator-entry\\ \\
$t_0$&$=(t_{o,2} + t_{c,1})/2$\\
$l$&$=(x_2-x_1)/2$ \\
$\epsilon$&$=(t_{o,2} - t_{c,1})/2$\\  \\
$L_i, L_j, L_k$&$=x_d-x_i$ etc\\
$L$&$=(L_1+L_2)/2$\\ \\
$\lambda_1$, $\lambda_2$ & wavelength boundaries of a given tof-channel\\ 
$\lambda_0$ & $=(\lambda_1+\lambda_2)/2$, nominal wavelength\\  \\
$r_1$, $r_2$&half-width of the 1st and 2nd slit of collimator\\
$d_c$&distance between the two slits of collimator\\
$\alpha_c$&$=(r_1-r_2)/d_c$, beam divergence due to collimation\\ 
$r$&radius of disks\\ \\
$\tau_p$& pulse half-width\\
$\tau_d$& half-width of tof-channel\\
$\tau_{\textrm{cr,}c}$, $\tau_{\textrm{cr,}o}$& crossing-time for beam closure and opening\\ \\
$\theta_0$& nominal specular reflection angle (tilt of the sample)\\
$\theta$& specular reflection angle\\
$\alpha$&deviation of neutrons due to beam divergence\\
$\gamma$& deviation due to gravity at the sample position\\ \\
$D(\lambda)$& wavelength transmission probability density\\
$H_B(\lambda)$& wavelength distribution of incident beam \\
$H(\lambda)$& wavelength distribution on the sample\\
$P(\alpha)$&anglular distribution due to beam divergence\\
$J(\gamma)$&anglular distribution on the sample due to gravity\\
$G(\theta)$&distribution of incidence angle\\
$q$&$={4\pi\sin(\theta)}/\lambda$, transfer vector\\
$R(q)$& resolution function for transfer vector\\ \\
$m(q)$& theoretical reflectivity curve (model)\\
$M\subscr{th}(q)$& theoretical measurement of $m(q)$ accounting for $R(q)$\\
$M$&actual measurement of $m$
\end{tabular}
\end{center}
\label{tab2}
\end{table}

\subsection{Wavelength transmission function}

For subsequent calculations of the exact resolution function, it is needed to consider probability density functions rather than their half-widths. The previous section can be formalized as follows. Let us consider the probability for a neutron of wavelength $\lambda= h_m/v$ to reach the detector at time $t_d$. The probability that this neutron passed through the first disk before it closed is the unit step function $f_1(\lambda)=[t_d-L_1/v<t_{c,1} ]=[(t_d-t_{c,1})/L_1<1/v]$, where $[...]=1$ if the condition inside the brackets is true and 0 if not.
The probability to pass the second disk after it opens is the unit step function $f_2(\lambda)=[t_{o,2}<t_d-L_2/v]=[1/v<(t_d-t_{o,2})/L_2]$. This rewrites as\,:
\begin{equation}
\begin{array}{l}
f_1(\lambda)=[\lambda_1<\lambda ]\\
f_2(\lambda)=[\lambda < \lambda_2]
\end{array}
\end{equation}
The probability density $D_0(\lambda)$ a neutron of wavelength $\lambda$ reaches the detector at time $t_d$ is the probability to fulfill these two independent conditions.  To a normalization factor one gets\,:
\begin{equation}\label{D0}
D_0(\lambda) = f_1(\lambda)f_2(\lambda) =[\lambda_1 < \lambda < \lambda_2]
\end{equation}

\subsection{Three-disk chopper}

\begin{figure}
\centering
\includegraphics[width=\wb\linewidth]{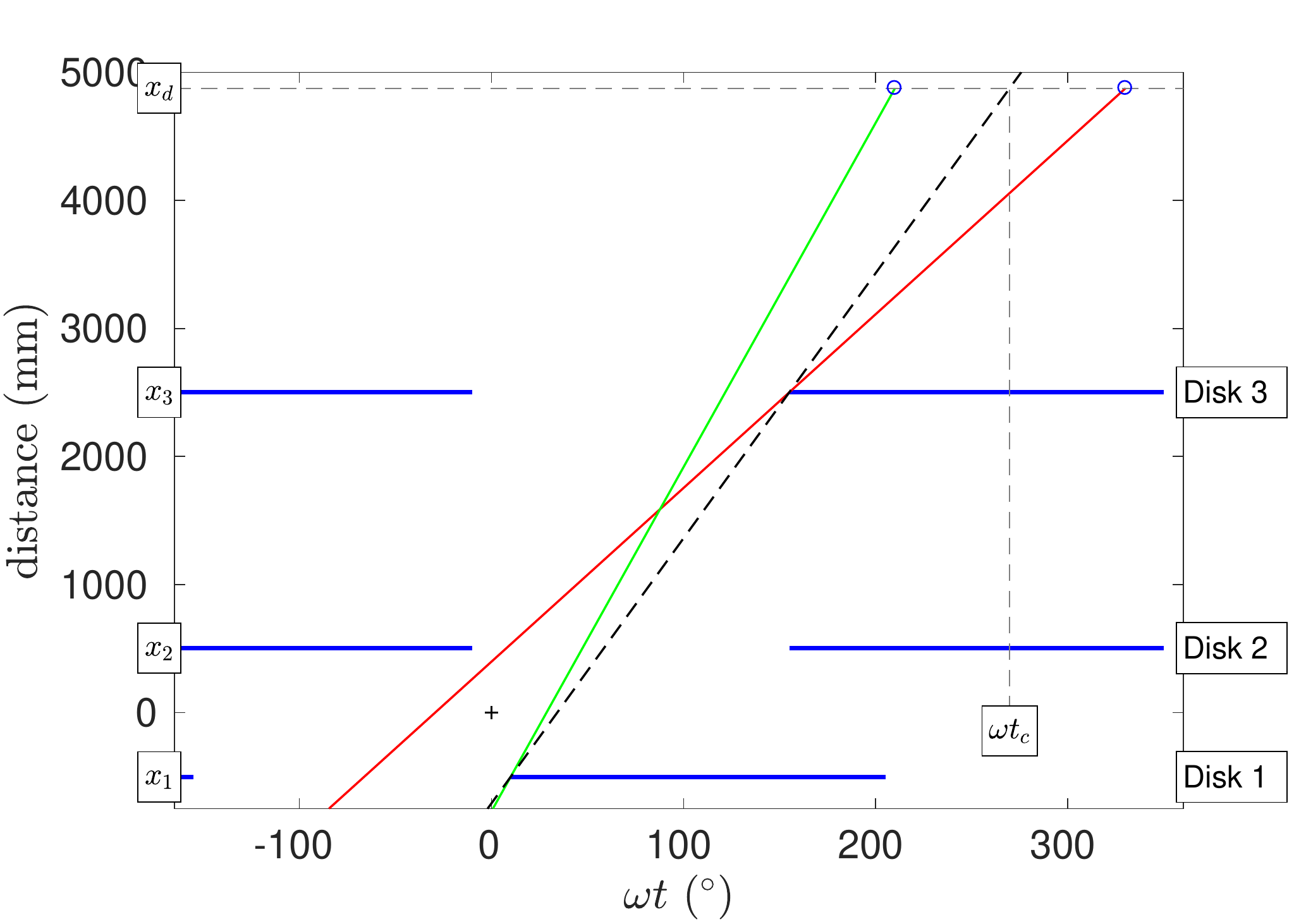} 
\caption{Three-disk chopper: compared to Fig.\ref{chopper1}, the 3rd disk comes into play for long time-of-flight. The dashed line in black with slope $(x_3-x_1)/(\varphi_{c,3}-\varphi_{c,1})=(x_d-x_1)/(\omega t_c-\varphi_{c,1})$ delimits two regimes for the velocity $v_1$ of fastest neutrons of one given tof-channel\,: $v_1$ is limited either by the closure of disk~1 (green), or by the one of disk~3 (red).  The same occurs for slow neutrons of velocity $v_2$ (case not shown).}
\label{chopper2}
\end{figure}

The general case of a three-disk chopper is a bit more complicated because the opening and closure in Eq.\ref{D0} that chops the neutron-beam can come from any pair of disks. For instance in Fig.\ref{chopper2}, the velocity $v_1$ of fastest neutrons reaching the detector at time $t_d$ is limited by the closure either of disk~1 (at short time $t_d$) or of disk~3 (at long time $t_d$). This can be formalized as follows. For each disk $k$ let us  denote\,:
\begin{equation}\label{vi}
v_{c,k}^{-1}=\max\left\{{0,\frac{t_d-t_{c,k}}{L_k}}\right\}
\quad\textrm{and}\quad
v_{o,k}^{-1}=\max\left\{{0,\frac{t_d-t_{o,k}}{L_k}}\right\}
\end{equation}
Let us define $i$, $j$ the indexes such as\,:
\begin{equation}\label{index}
\displaystyle v_{c,i}^{-1}=\max_{k\in\{1, 2, 3\}}\{v_{c,k}^{-1}\}
\quad\textrm{and}\quad
\displaystyle v_{o,j}^{-1}=\min_{k\in\{1, 2, 3\}}\{v_{o,k}^{-1}\}
\end{equation}
Then, all expressions of the previous section can be generalized by replacing $(1,2)$ by $(i,j)$.
Note that in a standard configuration $t_{c,1}\simeq t_{o,2}\simeq t_{o,3}$ so that the third disk comes into play only for long detection time. In this case, for most of time-of-flight channels  $(i,j)=(1,2)$ that corresponds to the double-disk regime. However this is not general and $(i,j)$, which depends on the relative phases of the three disks, depends also on the time-of-flight channel. 

\subsection{Crossing times and width of time-of-flight channel}

The previous section implicitly considers a neutron beam of zero width.
Due to its finite size, the beam aperture (and closure) due to the passage of a disk is not instantaneous and not correctely descibed by a unit step function\,\cite{nelson_towards_2013}. If the collimation is done with rectangular slits parallel to the edge of the disk windows, the transmitted beam area varies linearly with time, from 0 to its maximum value in a time $2\tau\subscr{cr}=2r_a/\omega r$, with $r_a$ the half-width of the beam and $r$ the radius of chopper disks. Actually, $r_a$ depends on the beam divergence due to collimation (see section \ref{beamdiv}) and thus to the position of the disk with respect of the collimator. Let us denote, $r_1$ and $r_2$ the half-width of the two slits of the collimator, $d_c$ their distance and $\alpha_c=(r_1-r_2)/d_c$. Then the half-width of the beam at position $x$ is\,:
\begin{equation}\label{beamsize}
r_a(x)=r_1+\alpha_c(x\subscr{in}-x)
\end{equation}
where $x\subscr{in}$ is the position of the collimator entry. For $x\simeq x\subscr{in}$, $r_a(x)\simeq r_1$\,\cite{nelson_towards_2013} but even if this condition is fulfilled of one disk of the pair $(i,j)$, it may not be for the other because their positions are different. Thus, except in case of a specially small beam divergence, the beam size is different for the two disks. Also, typical values for $r_1$,  $r_2$ (Table \ref{tab1}) and $r=300$\,mm, give $\omega \tau\subscr{cr} $ between 0.1 and 0.6$^\circ$, which is not negligible compared to the pulse half-width $\omega\tau_p$ at short wavelength $\lambda_0$. The general expression for the beam size is then preferred.
The corresponding crossing times for beam closure ($\tau_{\textrm{cr,}c}$) and opening ($ \tau_{\textrm{cr,}o}$) are\,: 
\begin{equation}\label{tcross}
 \tau_{\textrm{cr,}c}=r_a(x_i)/\omega r\quad\textrm{and}\quad \tau_{\textrm{cr,}o}=r_a(x_j)/\omega r
\end{equation}
The unit step functions $f_1$ and $f_2$ for the probabilities to pass the first and second disk in Eq.\ref{D0} have to be replaced by two linear piecewise functions $g_1$ and $g_2$ corresponding to the linear variations of the corresponding beam areas between 0 and maxima\,:
\begin{equation}
\begin{array}{c}
g_1(\lambda)=\left\{{\begin{array}{ll}
1& t_d<t_c-\tau_{\textrm{cr,}c}\\
\frac{1}{2}+(t_c-t_d)/2\tau_{\textrm{cr,}c}& t_c-\tau_{\textrm{cr,}c}\le t_d\le t_c+\tau_{\textrm{cr,}c}\\
0 & t_c+\tau_{\textrm{cr,}c}<t_d
\end{array}
}\right.\\ \\
g_2(\lambda)=\left\{{\begin{array}{ll}
0& t_d<t_c-\tau_{\textrm{cr,}o}\\
\frac{1}{2}+(t_d-t_o)/2\tau_{\textrm{cr,}o}& t_o-\tau_{\textrm{cr,}o}\le t_d\le t_o+\tau_{\textrm{cr,}o}\\
1 & t_o+\tau_{\textrm{cr,}o}<t_d
\end{array}
}\right.\\ \\
\textrm{with}\quad t_c = t_{c,i}+L_i/v\quad\textrm{and}\quad t_o = t_{o,i}+L_j/v
\end{array}
\end{equation}
As a function of wavelength this can be rewritten in terms of convolution as\,:
\begin{equation}\label{g1g2}
\begin{array}{rl}
g_1(\lambda)&\displaystyle=f_1(\lambda)*\left[{-\frac{h_m\tau_{\textrm{cr,}c}}{L_i}<\lambda-\lambda_1<\frac{h_m\tau_{\textrm{cr,}c}}{L_i}}\right]\\ \\
g_2(\lambda)&\displaystyle=f_2(\lambda)*\left[{- \frac{h_m\tau_{\textrm{cr,}o}}{L_j}<\lambda-\lambda_2< \frac{h_m\tau_{\textrm{cr,}o}}{L_j}}\right]
\end{array}
\end{equation}
The probability density for the transmission $g_1(\lambda)g_2(\lambda)$ is a piecewise function with up to five pieces (instead of a boxcar function for $D_0(\lambda)$), which is not symmetrical (because $\tau_{\textrm{cr,}c}\ne\tau_{\textrm{cr,}o}$ and $L_i\ne L_j$) and not always linear (\textit{e.g.} for small $\lambda_2-\lambda_1$).

Neutron counters record the number of neutron-arrivals within the time interval $[t_d-\tau_d, t_d+\tau_d]$. For instance 256 time-of-fligth channels covering 360$^\circ$ of data acquisition gives $\omega\tau_d\simeq 0.7^\circ$, which is not always negligible compared to the pulse half-width $\omega\tau_p$, specially at short wavelength $\lambda_0$ and is comparable to the crossing time for beam closure and opening $\omega\tau_{\textrm{cr,}o}$ and $\omega\tau_{\textrm{cr,}c}$. In addition, note that in order to gain in counting-statistics channel-binning is often performed after data acquisition. This operation amounts to use larger channel width $\tau_d$. Similarly to the crossing time, the effect of the channel width is not symmetrical for the disk $i$ closure and disk $j$ aperture, because $L_i\ne L_j$. By taking into account the crossing times and channel width, one can finally write the transmission function $D(\lambda)$ as\,:
\begin{equation}\label{D}
\begin{array}{c}
D(\lambda)=h_1(\lambda)h_2(\lambda)\\ \\
\textrm{with}\left\{{
\begin{array}{rl}
h_1(\lambda)&\displaystyle=g_1(\lambda)*\left[{-\frac{h_m\tau_d}{L_i}<\lambda-\lambda_1<\frac{h_m\tau_d}{L_i}}\right]\\ \\
h_2(\lambda)&\displaystyle=g_2(\lambda)*\left[{- \frac{h_m\tau_d}{L_j}<\lambda-\lambda_2< \frac{h_m\tau_d}{L_j}}\right]
\end{array}}\right.
\end{array}
\end{equation}
This results again in a piecewise  function that would be a little tedious to write explicitly, because it is made of up to seven pieces with different cases to consider depending on the order in which are $\tau_p$, $\tau_d$, $\tau_{\textrm{cr,}c}$, $\tau_{\textrm{cr,}o}$. Numerically, Eq.\ref{D} can be computed as is  (see python code \url{https://bitbucket.org/LLBhermes/pytof/}).

\begin{figure}
\centering
\includegraphics[width=\wb\linewidth]{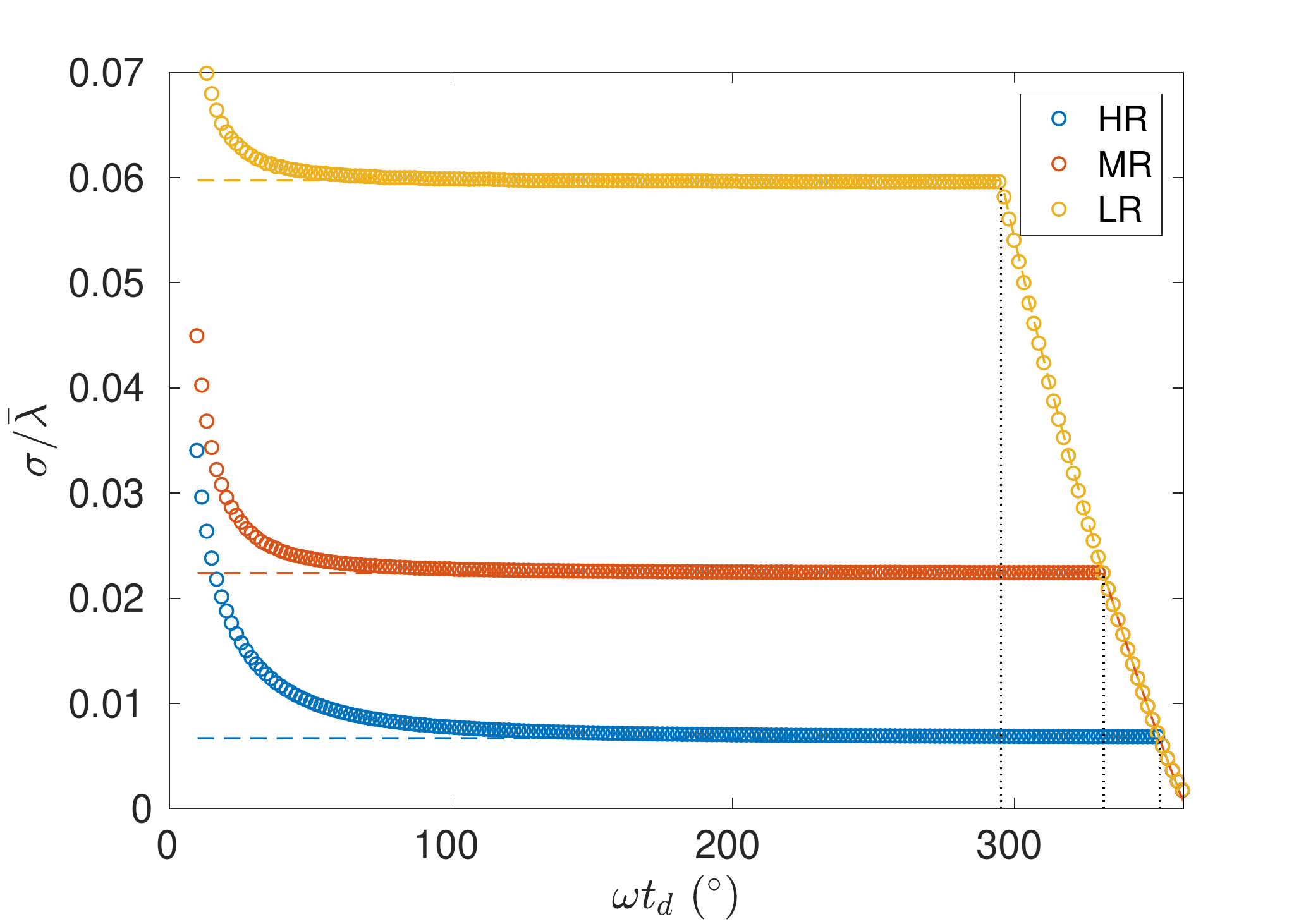} 
\caption{Ratio of standard deviation $\sigma$ of $D(\lambda)$ to mean value $\bar\lambda$ vs. phase of tof-channels for the three typical configurations described in Table \ref{tab1}. The dotted vertical lines mark the separation $\omega t_c$ of the two regimes delimited by the dashed line in Fig.\ref{chopper2}.}
\label{relstd}
\end{figure}

\begin{figure}\label{figBin}
\centering
\includegraphics[width=\wa\linewidth]{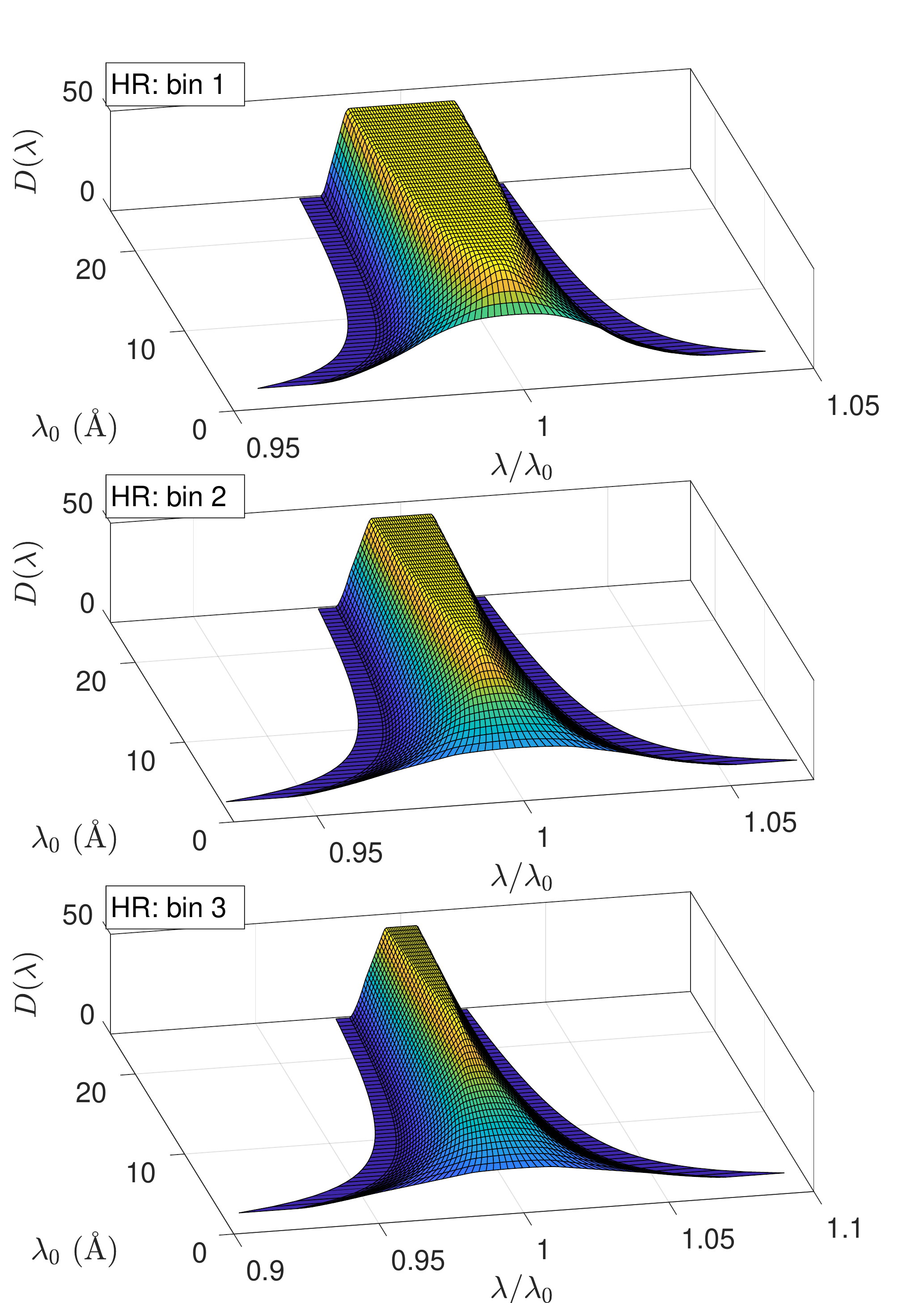}
\caption{Wavelength distributions for the chopper transmission $D(\lambda)$ (Eq.\ref{D}) vs. $\lambda/\lambda_0$ (x-axis) vs. tof-channel at nominal wavelength $\lambda_0$ (y-axis). Calculated  for the high resolution configuration (HR) of Table\,\ref{tab1} with 300 (no channel binning), 150 (data binning by a factor 2) and 100 (data binning by a factor 3) tof-channels from 30$^\circ$ to 300$^\circ$ of a chopper revolution.}
\end{figure}

In Fig.\ref{relstd}, the ratio of standard deviation $\sigma$ of $D(\lambda)$ to its mean value $\bar\lambda$ is plotted for 256 time-of-flight channels covering 360$^\circ$ of a chopper revolution for the three typical configurations of Table \ref{tab1}. The curves show a plateau corresponding to the "double-disk" chopper regime described by Eq.\ref{epsilon0}. At large $\omega\tau_d$ (large $\lambda_0$), the decrease in resolution is due to the third disk that comes into play, whereas at small $\omega\tau_d$ (small $\lambda_0$) departure from the plateau is due to crossing-times and channel width.

The effect of channel binning on $D(\lambda)$ is plotted in Fig.\ref{figBin} for the high-resolution configuration of Table\,\ref{tab1}. Basically, $D(\lambda)$ continuously changes from a nearly boxcar profile at large wavelength $\lambda_0$ to a ``bell curve'' (but never gaussian) at small wavelength $\lambda_0$. The time-of-flight channels at which this change appears is shifted to high wavelength $\lambda_0$ as bin size increases.

\subsection{Phase calibration and tilt of the chopper axis}

The time resolved neutron counting is triggered at each revolution by the chopper electronics. On this subject, the only unknown parameter is the phase shift $\phi\subscr{trig}$ of this trigger with respect to the physical origin of phases. This point is resolved by measuring the transmission of a crystalline material (\textit{e.g.} a graphite crystal) that displays a characteristic attenuation at wavelength $\lambda^*$\,\cite{PhysRev.71.589}. No matter the value of $\lambda^*$, the corresponding time-of-flight $t^*$ is constant and the related phase writes $\phi^*=\omega t^* +\phi\subscr{trig}$. Measuring $\phi^*$ as a function of $\omega$ and extrapolating to $\omega = 0$ gives $\phi\subscr{trig}$.

Taking the rotation of disk~1 as reference, the phase shifts of disk~2 and 3 are chosen by the experimentalist and kept constant by the electronics with a control loop feedback, which ensures that no variation or drift occurs during measurements.
However, possible differences between phases setpoints and their actual values have to be measured. Another point to consider is that a vertical tilt or misalignment of the centers of the three disks with the spectrometer axis affects the kinematics line of neutrons allowed to pass through the chopper in the same manner as a phase difference. Hence the need of disk-phase calibration.

\begin{figure}
\centering
\includegraphics[width=\wb\linewidth]{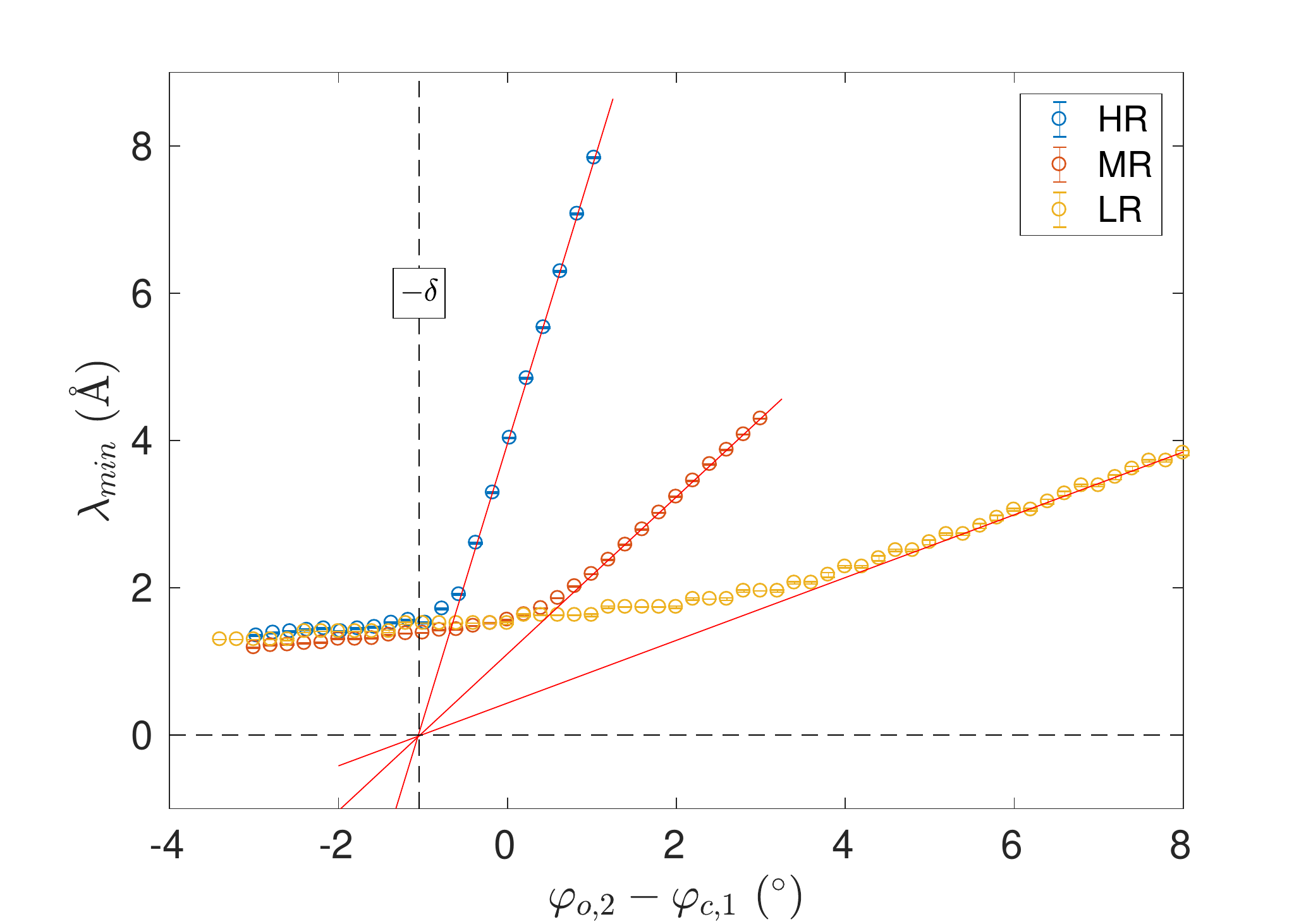} \\
\caption{Minimum wavelength $\lambda_{min}$ as a function of $\varphi_{o,2} - \varphi_{c,1}$, measured for the three different configurations of Table\,\ref{tab1}; at $\omega=30$\,Hz. Extrapolation to $\lambda_{min}=0$ gives the angle $\delta=(1\pm0.02)^{\circ}$ that has to be added to the phase of disk~2.}
\label{figmin}
\end{figure}

The phase of disk~2 affects the short-wavelength cut-off ($\lambda_1$) and the chopper resolution in the first regime of Fig.\ref{relstd}, whereas the phase of disk~3 affects the large-wavelength cutoff ($\lambda_2$) in the second regime and the time-of-flight channel at which this regime begins. Let us first consider the former.
From Fig.\ref{chopper1}, one can see that the fastest neutrons passing through the chopper are such that $\lambda_{min}/h_m=1/v_{max}=\epsilon/l$. 
Measuring the transmission of the chopper, $\lambda_{min}$ corresponds to the nominal wavelength of the tof-channel at which the spectrum departs from the background. As the spectrum is quite abrupt in this region (see Fig.\ref{figH}) this is very accurate. Of course, converting channel to wavelength assumes that we neglect the resolution function (because at this point, we do not know it yet). But the consequence of this approximation is negligible as $\Delta \lambda$ is very small at short wavelength. In Fig.\ref{figmin}, the measured $\lambda_{min}$ is plotted as a function of $\varphi_{o,2} - \varphi_{c,1}=2\omega\epsilon$, the linear behavior of the cut-off does not passes through the origin\,: a constant $\delta$ (here equal to 1$^\circ$ for the three configurations) should be added to $\varphi_{o,2} - \varphi_{c,1}$ in order to obtain the correct value for $\lambda_{min}$. 
In practice, the phases $\varphi_{o,2}$ and $\varphi_{c,2}$  of disk~2 in all equations will be replaced by $\varphi_{o,2}+\delta$ and $\varphi_{c,2}+\delta$ and the corresponding times $t_{o,2}$ and $t_{c,2}$ changed accordingly.
Note that this method for measuring $\varphi_{o,2}$ is equivalent to measuring the intensity of a monochromatic beam as a function of the phase $\varphi_{o,2} - \varphi_{c,1}$\,\cite{gutfreund_towards_2018}, but is simpler to perform routinely because it does not require setting up a monochromator on the beamline.

\begin{figure}
\centering
\includegraphics[width=\wb\linewidth]{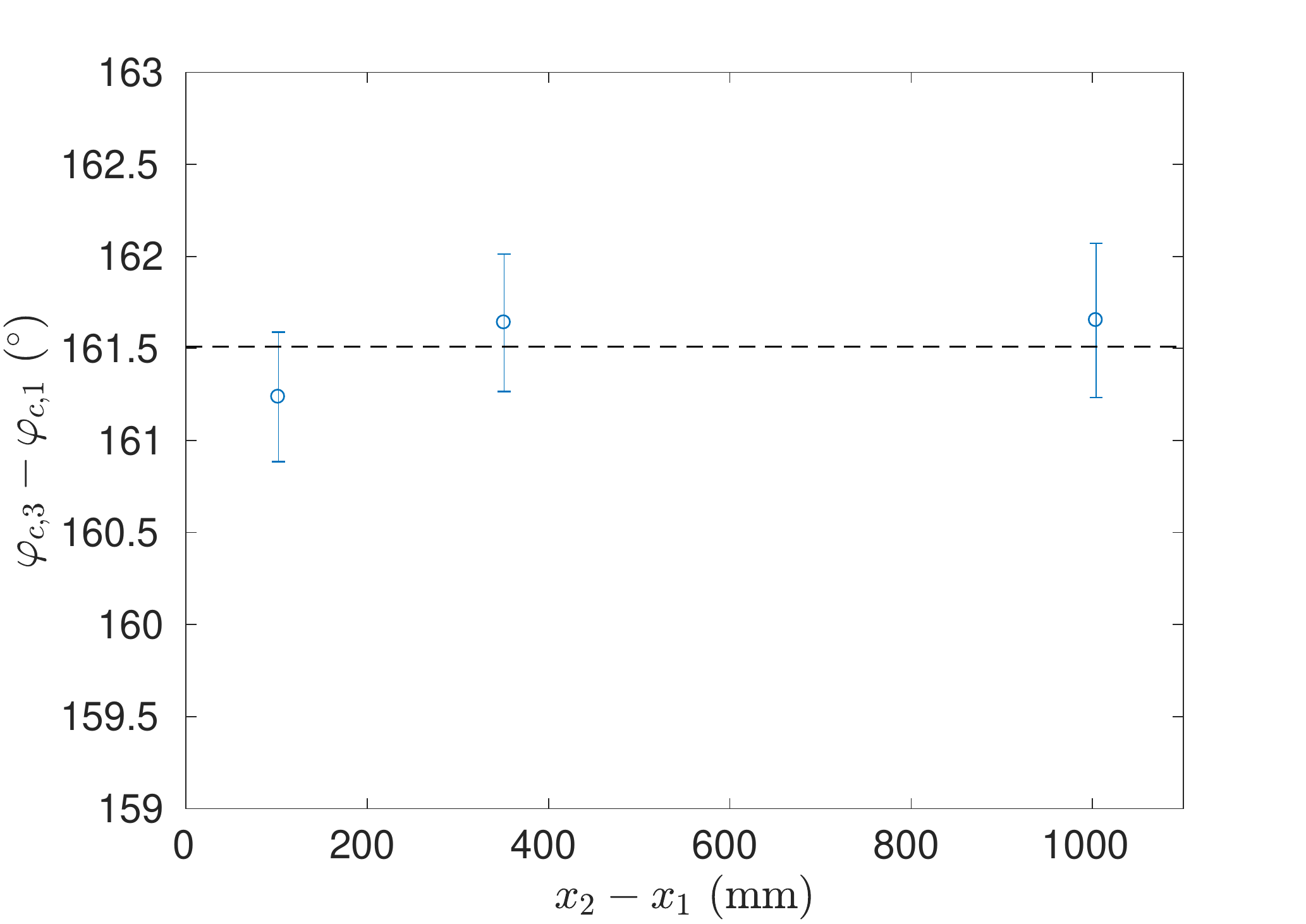} 
\caption{Value of the phase $\varphi_{c,3}$ deduced from Eq.\ref{cutoff} and the measurement of the phase $\omega t_c$ of the large wavelength cut-off for a setpoint $\varphi_{c,3}=165^\circ$. The difference with this setpoint gives $\eta=-(3.5\pm0.3)^{\circ}$ that has to be added to the phase of disk~3.}
\label{figmax}
\end{figure}

In the same way, the large-wavelength cut-off of the chopper transmission allows us to determine the actual phase of disk~3.  The phase $\omega t_c$ of the cut-off (see Fig.\ref {chopper2}), is such as\,:
\begin{equation}\label{cutoff}
\frac{\omega t_c - \varphi_{c,1}}{(x_d-x_1)/(x_3-x_1)}=\varphi_{c,3}
\end{equation}
From the measurement of $\omega t_c$ the actual value of $\varphi_{c,3}$ can be deduced (Fig.\ref{figmax}). Here, we found that a constant value $\eta=-3.5^\circ$ has to be added to the phase of disk~3. In the general case, the phase $\varphi_{o,3}$ and $\varphi_{c,3}$  of disk~3 in all equations will be replaced by $\varphi_{o,3}+\eta$ and $\varphi_{c,3}+\eta$  and the corresponding times $t_{o,3}$ and $t_{c,3}$ changed accordingly.

\subsection{Gravity effects on tilt-angles}\label{secgrav1}

Due to gravity, neutrons flight is parabolic. The deviation from a straight line and horizontal trajectory increases with the time-of-flight leading to actual values for the vertical tilt-angles $\delta$ and $\eta$ of the previous section that depend on wavelength.

Let us assume that the detector and the collimator slits are properly aligned to the neutron guide using a criterion of maximum neutron flux measured by integrating over the whole wavelength-distribution. As neutrons of short wavelength are majority (see Fig.\ref{figH} in the next section) and neglecting their deviation due to gravity, the line joining the two slits of the collimator is almost horizontal. The apex of parabolic trajectories of all neutrons is thus at the middle $x_c$ between the two slits of the collimator. Let us denote $v_x=h_m/\lambda$ and $v_z$ the horizontal and vertical components of neutrons velocity, respectively. At the apex, $v_z=0$. At any other position $x$, the deflection angle $\zeta$ is $\tan^{-1}({v_z}/{v_x})\simeq {v_z}/{v_x}$. The time-of-flight to cover the distance $|x-x_c|$ is $|x-x_c|/v_x$, leading to $v_z=-g(x-x_c)/v_x$, with $g$ the gravitational acceleration. Thus:
\begin{equation}\label{eqgamma1}
\begin{array}{c}
\displaystyle\zeta(x)=\frac{-g(x-x_c)}{v_x^2}=-(x-x_c) \frac{g}{h_m^2} \lambda^2\\ \\
\displaystyle\textrm{with }\quad {g}/{h_m^2}\simeq 6.27\times 10^{-7}\textrm{m}^{-1}\textrm{\AA}^{-2}
\end{array}
\end{equation}
Given two points (\textit{e.g.} the edges of the two chopper-disks $i$ and $j$) of the parabola, the tangent at the middle is parallel to the chord. Thus, the tilt angles $\delta$ and $\eta$  viewed by neutrons with parabolic trajectories are
\begin{equation}\label{deltaeta}
\begin{array}{c}
\displaystyle\delta(\lambda) = \delta(0) + \zeta\left({\frac{x_1+x_2}{2}}\right)\\ \\
\displaystyle\eta(\lambda)=\eta(0)+\zeta\left({\frac{x_1+x_3}{2}}\right)
\end{array}
\end{equation}
As these tilt-angles have an incidence on the boundaries $\lambda_1$ and $\lambda_2$ of each tof-channel, gravity should affect the wavelength resolution. 
However, this effect is very small. For instance, in the double-disk regime $(i, j)=(1, 2)$, only $\delta$ is relevant and increases $\lambda_2$ by $h_m \zeta((x_1+x_2)/2)/\omega L_2 $. 
For few meters long reflectometers, $\zeta((x_1+x_2)/2)$ is around 0.01$^\circ$ and $\lambda_2$ is shifted by around $10^{-3}\textrm{\AA}$, which is clearly negligible.

\subsection{Incident beam and detector efficiency}

\begin{figure}
\centering
\includegraphics[width=\wb\linewidth]{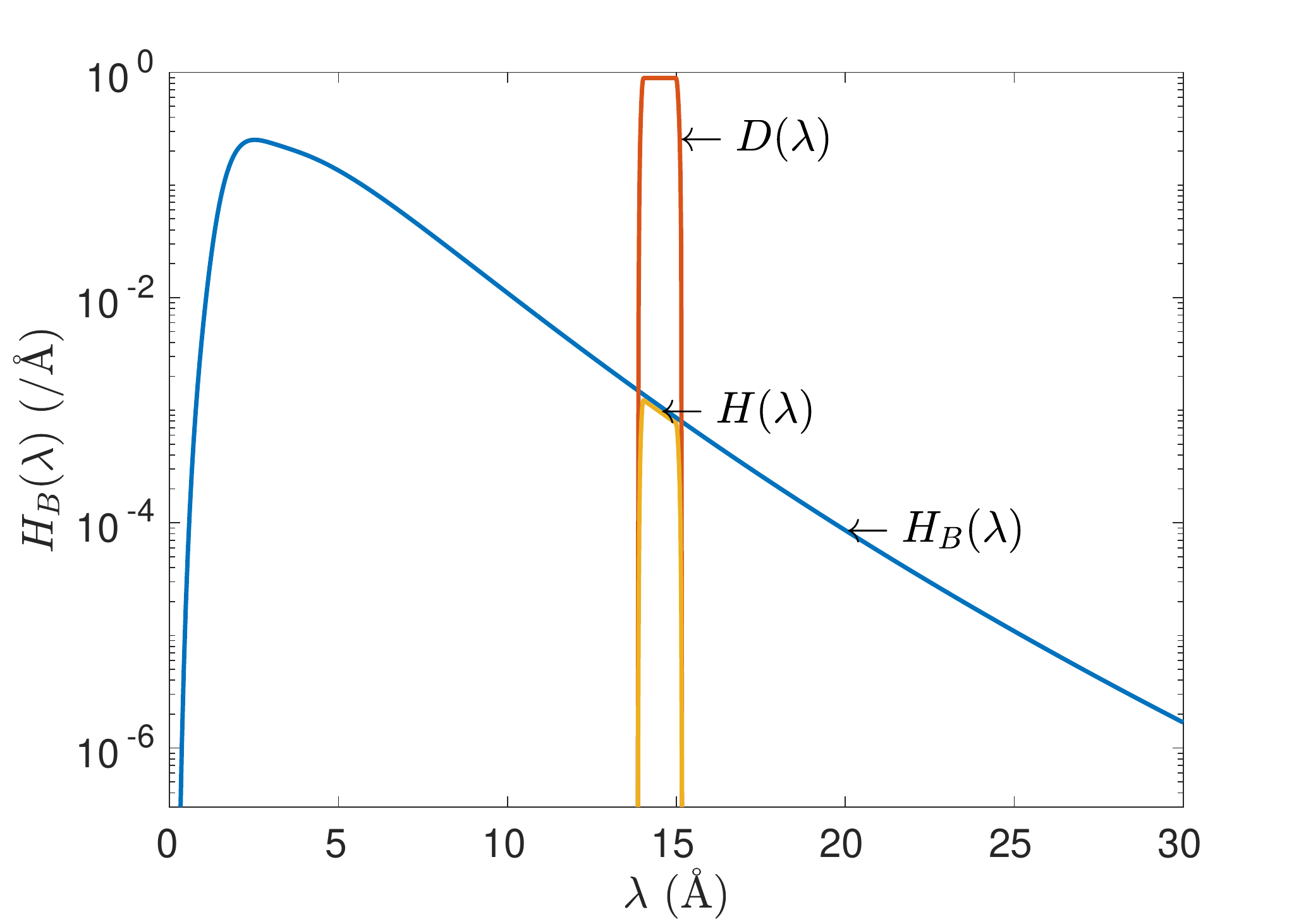} \\
\caption{Blue\,: effective wavelength distribution $H_B(\lambda)$ of the incident beam provided by the reactor Orph\'ee on the neutron guide G6-2. Red\,: transmission probability density $D(\lambda)$ for a tof-channel at 180$^\circ$ computed for the low resolution of Table\,\ref{tab1}. Orange\,: wavelength distributions $H(\lambda)$ (not normalized).}
\label{figH}
\end{figure}

Thermalized neutrons have a Maxwell-Boltzmann distribution of velocity that is altered by neutron guides. This distribution can not be directly measured but only its product $H_B(\lambda)$ with the detector efficiency that decays exponentially with wavelength.
For a given time-of-flight channel, chopping a neutron beam amounts to applying the transmission $D(\lambda)$ of Eq.\ref{D} to $H_B(\lambda)$\,\cite{nelson_towards_2013} (see Fig.\ref{figH}). The probability density of wavelength for neutrons recorded in this channel is
\begin{equation}\label{eqH}
H(\lambda)=H_B(\lambda)  D(\lambda)
\end{equation}
\noindent $H_B(\lambda)$ can be measured using a single-disk chopper with a small $\Delta\lambda$ (\textit{i.e.} much smaller than the three-disk chopper) independent of the time-of-flight channel. For further calculations, $H_B(\lambda)$ can then be properly parametrized using an \textit{ad hoc} function.

\begin{figure}
\centering
\includegraphics[width=\wa\linewidth]{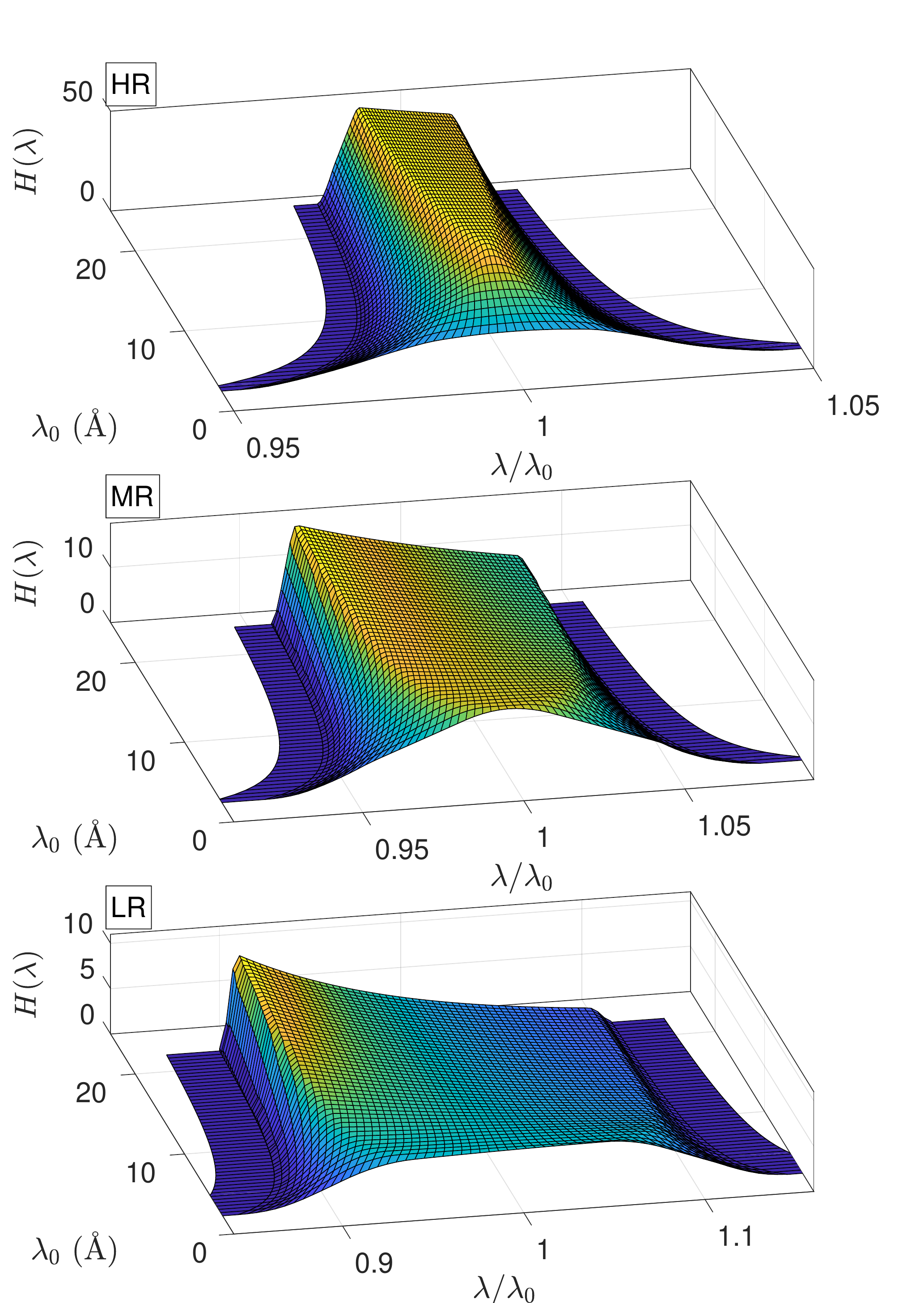}
\caption{Wavelength distributions $H(\lambda)$ (Eq.\ref{eqH}) vs. $\lambda/\lambda_0$ (x-axis) vs. tof-channel at nominal wavelength $\lambda_0$ (y-axis) for the three configurations of Table\,\ref{tab1}.}
\label{figH2}
\end{figure}

Note that measuring $H_B(\lambda)$ (which includes the detector efficiency) and using Eq.\ref{eqH}, is a way to account for the wavelength dependence of the detection-time mentioned in the literature\,\cite{gutfreund_towards_2018}.

\begin{figure}
\centering
\includegraphics[width=\wb\linewidth]{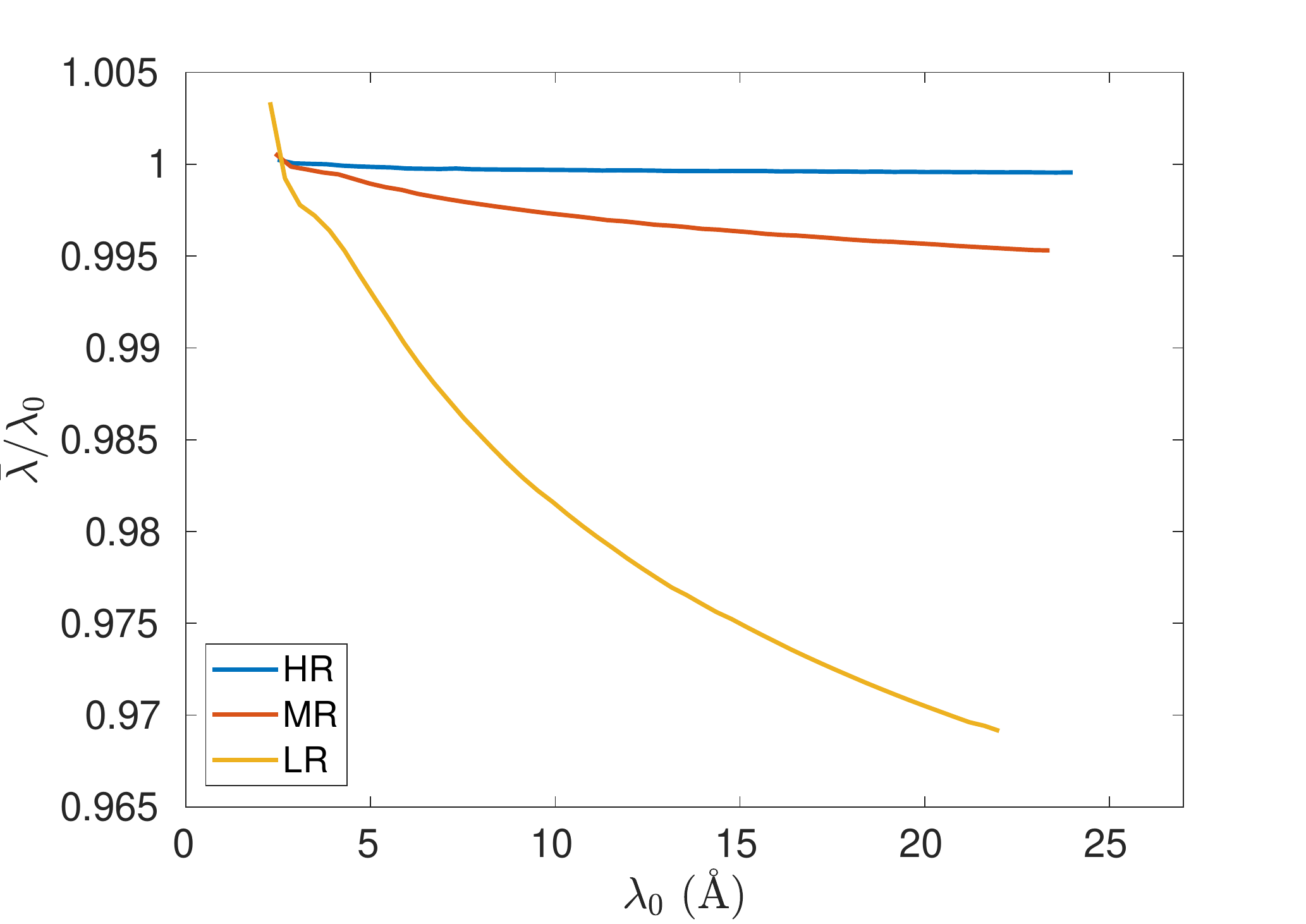}
\caption{Ratio of the mean wavelength $\bar\lambda$ of the distribution $H(\lambda)$ to the nominal wavelength  $\lambda_0$ vs. $\lambda_0$  for the three configurations of Table\,\ref{tab1}.}
\label{figl}
\end{figure}

In Fig.\ref{figH2}, the wavelength distribution $H$ is plotted for the different time-of-flight channels and for the three configurations of Table\,\ref{tab1}. The strong asymmetry of $H(\lambda)$ results in a mean wavelength $\bar\lambda=\int \lambda H(\lambda) \diff\lambda$ significantly different from $\lambda_0$. 
In order to render more clearly this difference, we plotted in Fig.\ref{figl} the ratio $\bar\lambda/\lambda_0$ as a function of the nominal wavelength $\lambda_0$ for the three different chopper configurations of Table\,\ref{tab1}. The wider the resolution, the more $\bar\lambda$ deviates from $\lambda_0$.

\section{Angular resolution}

In this section, we focus on beam collimation and gravity, which  both contribute to the final distribution of incidence angle of neutrons on the sample.

\subsection{Beam divergence}\label{beamdiv}

\begin{figure}
\centering
\includegraphics[width=\wb\linewidth]{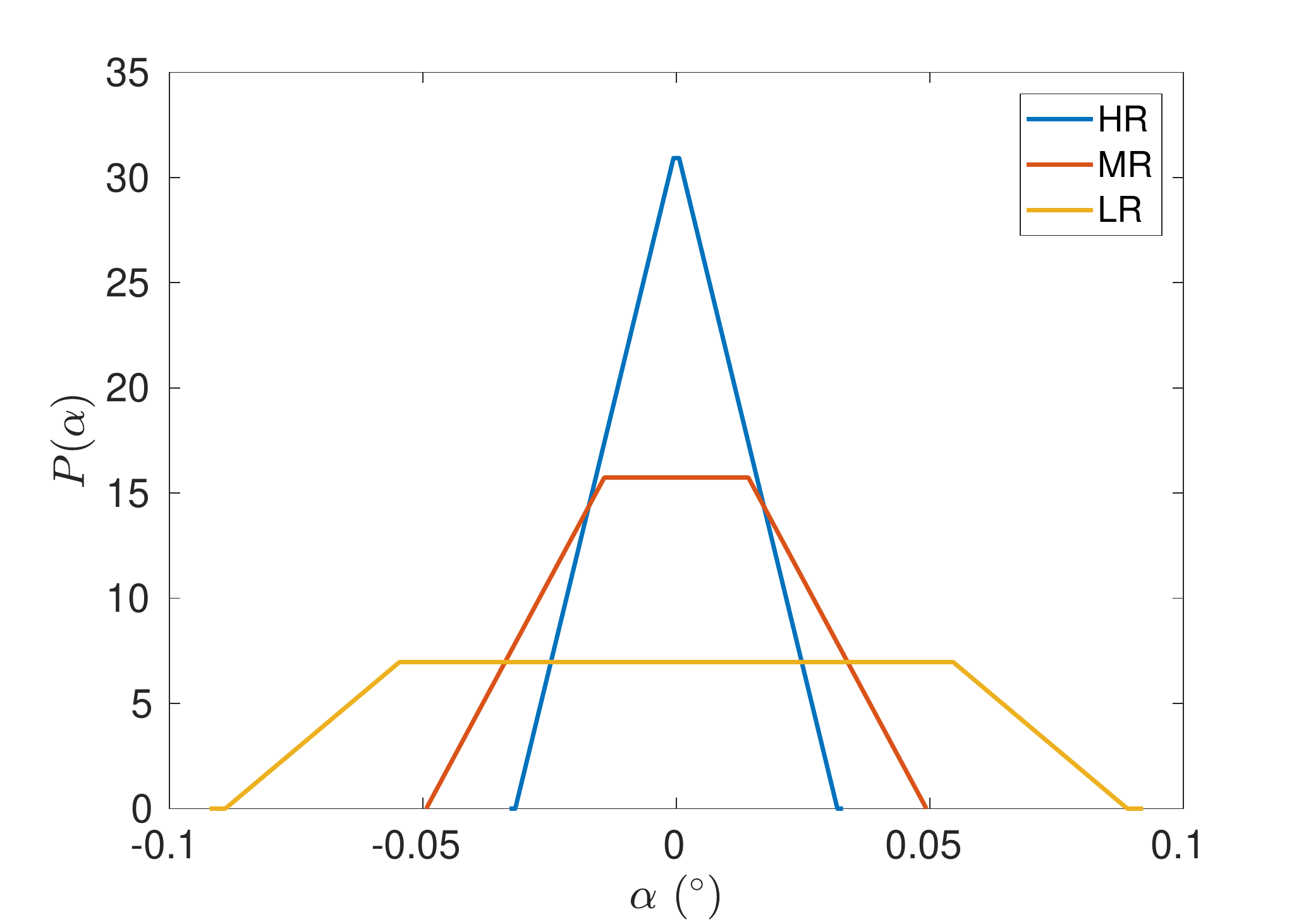}
\caption{Typical angular distribution $P(\alpha)$ due to beam divergence for the three standard collimations of Table\,\ref{tab1}.}
\label{figP}
\end{figure}

Let us consider the divergence in the vertical plane of a neutron-beam collimated with two horizontal slits of half-width $r_1$ and $r_2$, respectively, spaced by $d_c$. The distribution function $P(\alpha )$ of the angle of neutron trajectory results from the convolution of two boxcar functions centered on 0 and half-width $r_1/d_c$ and $r_2/d_c$, respectively\,:
\begin{equation}\label{eqP}
P(\alpha)=[-r_1/d_c<\alpha<r_1/d_c]*[-r_2/d_c<\alpha<r_2/d_c]
\end{equation}
This results in a symetrical linear piecewise function.
As convolution is commutative, the two boxcar functions are fully interchangeable and also the order of the values for $r_1$ and $r_2$ as far as only $P(\alpha)$ is concerned.
However for some reasons related to the beam size (\textit{e.g.} it is desired that the ``footprint'' $2r_2\sin(\theta)$ be smaller than the sample-size and also $r_2$ restrains the accuracy of the alignment of the sample), $r_1>r_2$ is preferred and $r_2$ is kept constant. In Fig.\ref{figP} typical curves for $P(\alpha)$ are plotted for standard collimations of Table\,\ref{tab1}.

The angular distribution $P(\alpha)$ so calculated implicitly assumes that the footprint of the beam on the sample (part of the sample illuminated by the beam) is smaller than the size $r_s$ of the sample in the $x$-direction. In case of small sample and very small tilt-angle $\theta_0$ of the sample with respect to the beam, it may happen that this condition is not fulfilled\,: $2r_2>r_s\sin(\theta_0)$. Then, $P(\alpha)$ has to be calculated only for the part of the beam that reachs the sample.
Eq.\ref{eqP} can account for this ``over illumination'' replacing $r_2$ by $r_s\sin(\theta_0)/2$ and $d_c$ by the distance between the first slit and the sample.

\subsection{Deviation due to gravity}\label{secgrav2}

As mentioned in section \ref{secgrav1}, gravity causes deviations of neutrons trajectories. The angle of incidence of neutrons reaching the sample depends on wavelength, which has to be accounted for in case specular reflectivity is measured in the vertical plane\,\cite{bodnarchuk_effect_2011}.

\begin{figure}
\centering
\includegraphics[width=\wb\linewidth]{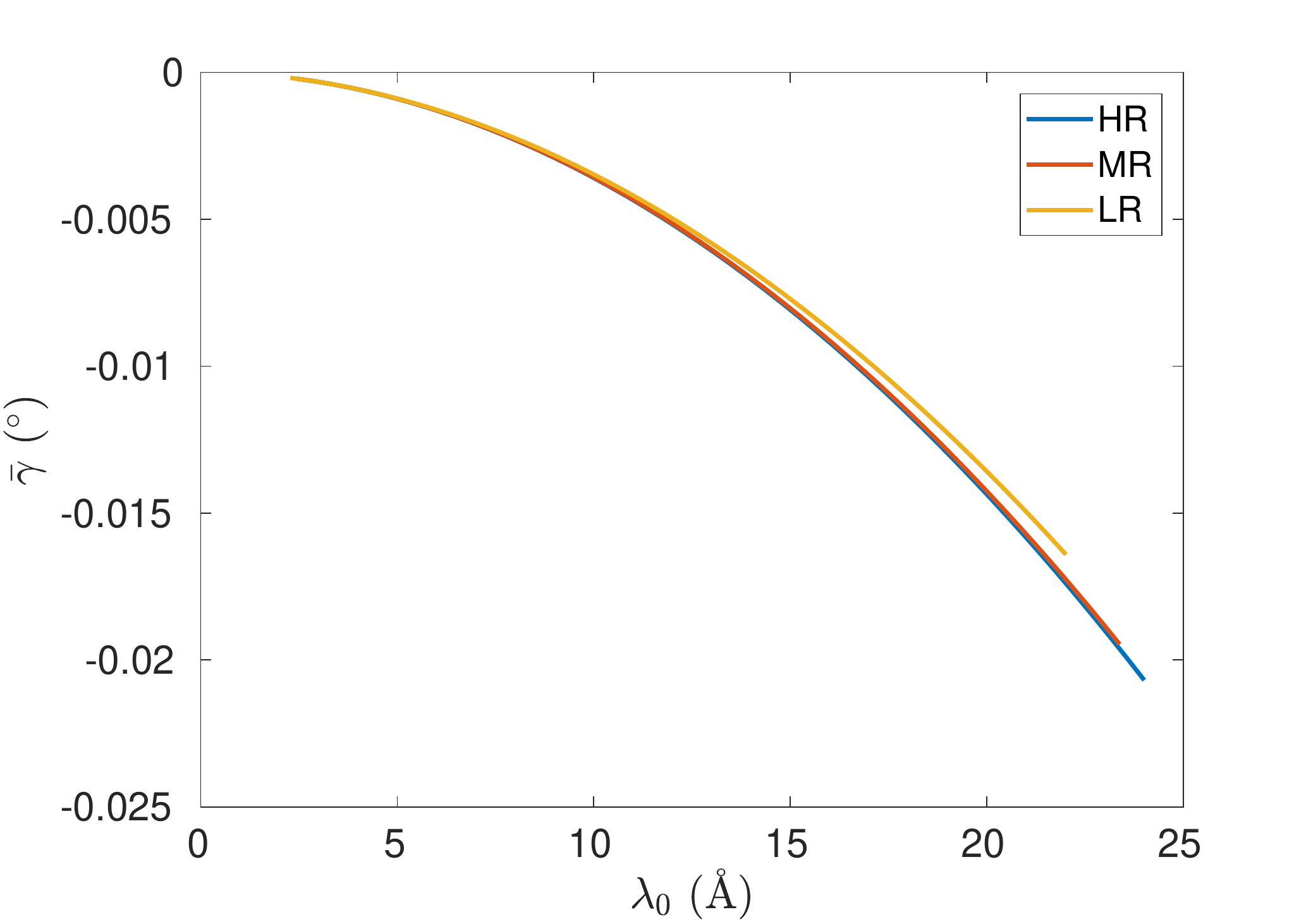}
\caption{Mean deviation $\bar\gamma$ of neutrons due to gravity for $x_s-x_c=1$\,m vs. $\lambda_0$ for the three configurations of Table.\ref{tab1}.}
\label{figgbar}
\end{figure}

If $x_s$ is the position of the sample, from Eq.\ref{eqgamma1} the deviation angle (which is always negative) on the sample is:
\begin{equation}\label{eqgamma2}
\displaystyle\gamma=\frac{-g (x_s-x_c)}{v_x^2}=-(c \lambda)^2
\quad\textrm{with }\quad c=\frac{(g(x_s-x_c))^{1/2}}{h_m}
\end{equation}
The distribution $J(\gamma)$ of the deviation is related to the distribution of wavelength $H(\lambda)$ via the general relation $J(\gamma)=H(\lambda)\times\mid\diff\lambda/\diff\gamma\mid$, with $\lambda=(-\gamma)^{1/2}/c$ (the inverse relation of Eq.\ref{eqgamma2}). One obtains\,:
\begin{equation}\label{eqJ}
J(\gamma)=\frac{1}{2c(-\gamma)^{1/2}} H((-\gamma)^{1/2}/c)
\end{equation}
Basically, $J(\gamma)$ has a shape comparable to $H(\lambda)$. Let us denote $[\gamma_1, \gamma_2]$ the support interval of $J(\gamma)$ and $\Delta\gamma=(\gamma_2-\gamma_1)/2$. From Eq.\ref{eqgamma2}, one obtains $\Delta\gamma=2(c\lambda_0)^2\times(\Delta\lambda/\lambda_0)$. The quantity $c$ is typically of the order of $8\times 10^{-4}\textrm{\AA}^{-1}$ for $x_s-x_c=1$\,m, thus in a standard configuration $\Delta\gamma$ is small compared to the beam divergence (support of $P(\alpha)$). However, in case the  relative angle resolution is very small compared to the wavelength resolution $H(\lambda)$, for instance in case of tof-channels binning to increase statistics\,\cite{Cubitt:ge5018},  or in case of narrow collimation due to small size of the sample etc., then gravity should affect the width of the angular resolution. In any case, the main effect of gravity is due to the mean value $\bar\gamma=\int \gamma J(\gamma) \diff\gamma$ that is non-zero and depends on the time-of-flight channel (\textit{i.e.} on $\lambda_0$). In Fig.\ref{figgbar}, the mean deviation $\bar\gamma$ is plotted as a function of $\lambda_0$.

Note that in case of over illumination of the sample (as discussed at the end of section \ref{beamdiv}), the middle of the collimator has to be understood as being in between the first slit and the sample.

\subsection{Distribution of incidence angle}

\begin{figure}
\centering
\includegraphics[width=\wa\linewidth]{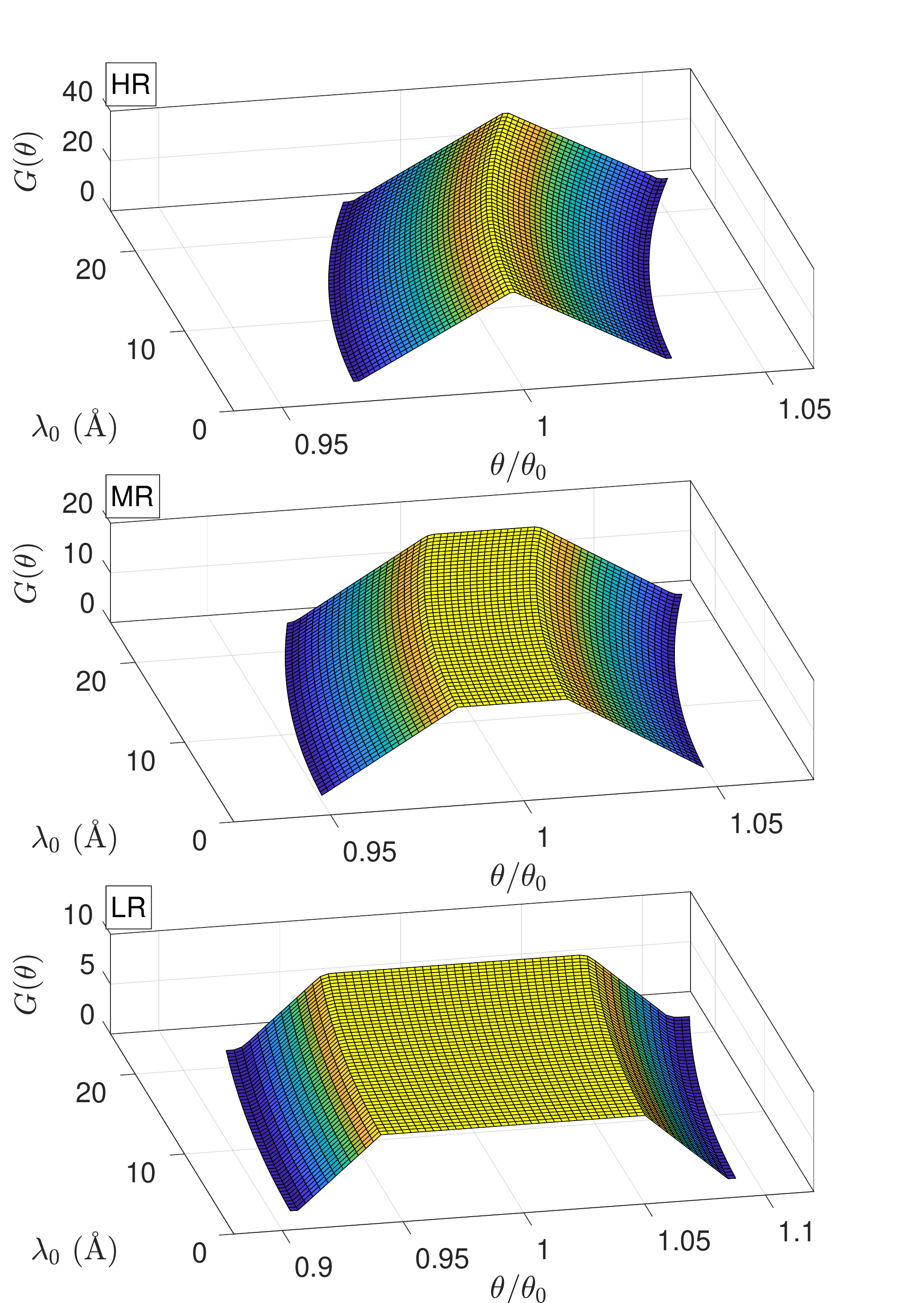}
\caption{Angular resolution $G(\theta)$  vs. $\theta/\theta_0$ (x-axis, where $\theta_0=1^\circ$ is the nominal angle) and vs. tof-channel at nominal wavelength $\lambda_0$ (y-axis). Calculated  for the three configurations of Table\,\ref{tab1}.}
\label{figG2}
\end{figure}

Let us denote $\theta_0$ the nominal angle (\textit{i.e.} the angle calculated from the inclination of the sample) of the neutron beam with respect to the interface under study. This angle is negative in case neutrons come from below  the interface and positive if they come from over (this sign is imposed by gravity). The actual incidence angle $\theta$ of neutrons on the sample is
\begin{equation}\label{eqtheta}
\theta=\theta_0 + \alpha -\gamma
\end{equation}
The distribution for $\theta$ is thus given by
\begin{equation}\label{eqG}
G(\theta)= \int P(\theta-\theta_0+\gamma)  J(\gamma) \diff \gamma
\end{equation}
As $J(\gamma)$ is much narrower than $P(\alpha)$ its shape affects only slightly the one of $G(\theta)$ but mainly induces an increase in the mean value $\bar\theta$ compared to the nominal value $\theta_0$. Note that as $\gamma$ is negative and due to the sign in Eq.\ref{eqtheta}, gravity increases $\mid\theta\mid$ for $\theta_0>0$ but decreases $\mid\theta\mid$ for $\theta_0<0$.
In Fig.\ref{figG2}, the angular resolution function $G(\theta)$ is plotted for $\theta_0=1^\circ$ for the different time-of-flight channels and the three typical configurations of the reflectometer. 

Note that $G(\theta)$ is plotted here for didactic reasons but is not needed for the calculation of the overall resolution. Only its elementary components $P$ and $J$ are used in the next section.

\section{Resolution of transfer vector}

The physical parameter related to structural informations on the measured samples is the transfer vector $q$ (the conjugate variable of distance) that is defined as
\begin{equation}\label{defq}
q={4\pi\sin(\theta)}/\lambda
\end{equation}
However, the only directly adjustable parameters of the spectrometer are $\theta$ and $\lambda$, both being distributed around $\theta_0$ and $\lambda_0$. For a given time-of-flight channel, the resolution function $R(q)$ of the transfer vector should be thus a generalized convolution of the probability densities $H(\lambda)$ (Fig.\ref{figH2}) and  $G(\theta)$  (Fig.\ref{figG2}) which corresponding random variables are combined following Eq.\ref{defq}. However due to gravity, $\theta$ and $\lambda$ are not strictly independent and convolving directly their densities is not correct even if it is a good approximation in standard configurations (because $J(\gamma)$ is narrow compared to $P(\alpha)$ in most cases). As moreover this approximation does not save computation time, we prefer an exact treatment that will remain valid even when the width of $J$ becomes significant (see section \ref{secgrav2}).

\begin{figure}
\centering
\includegraphics[width=\wa\linewidth]{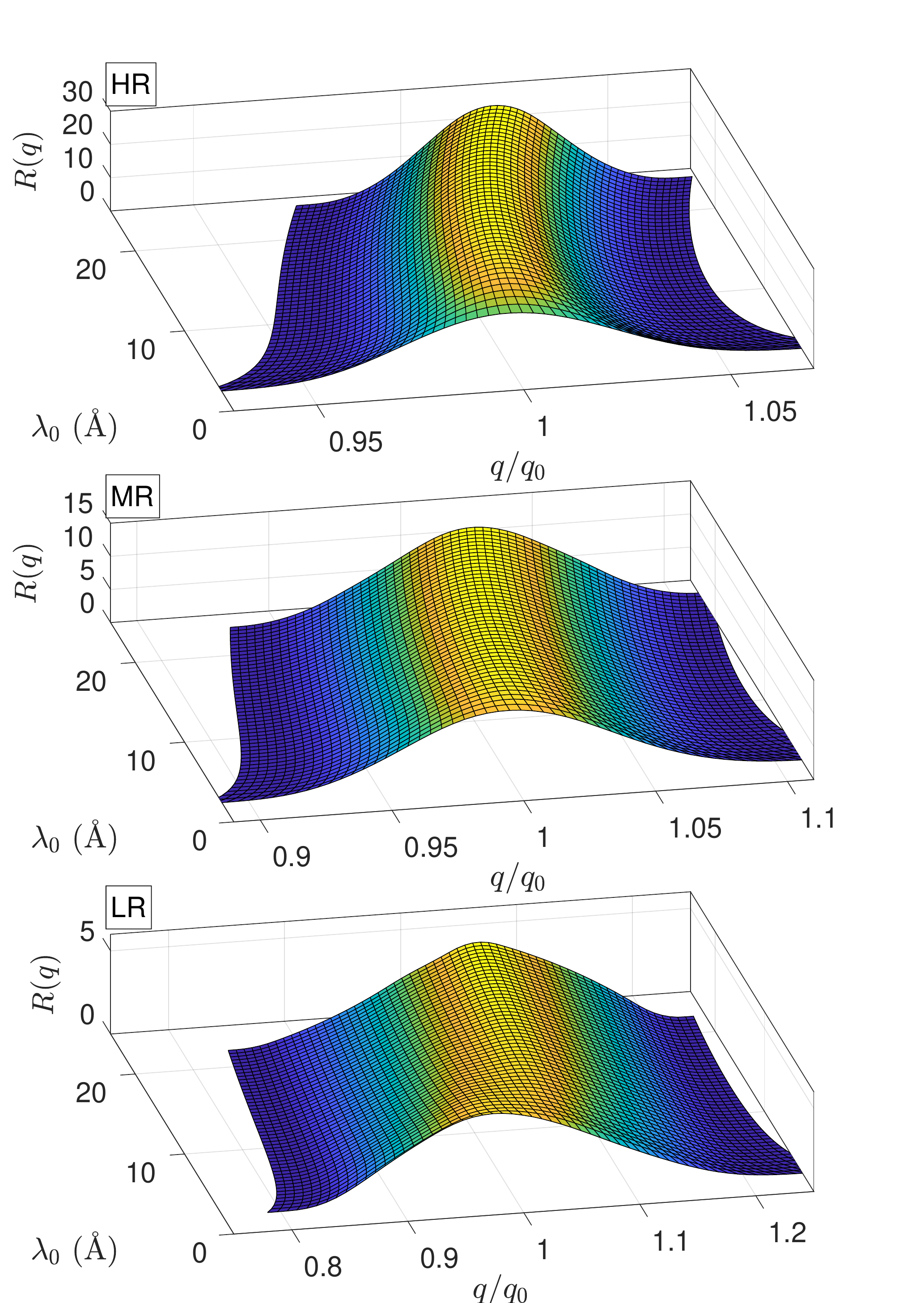}
\caption{Total resolution $R(q)$ of transfer vector $q$ vs. $q/q_0$ (x-axis, where $q_0=4\pi\sin(\theta_0)/\lambda_0$, $\theta_0=1^\circ$ is the nominal angle) vs. tof-channel at nominal wavelength $\lambda_0$ (y-axis). Calculated  for the three configurations of Table\,\ref{tab1}.}
\label{figR}
\end{figure}

\begin{figure}
\centering
\includegraphics[width=\wb\linewidth]{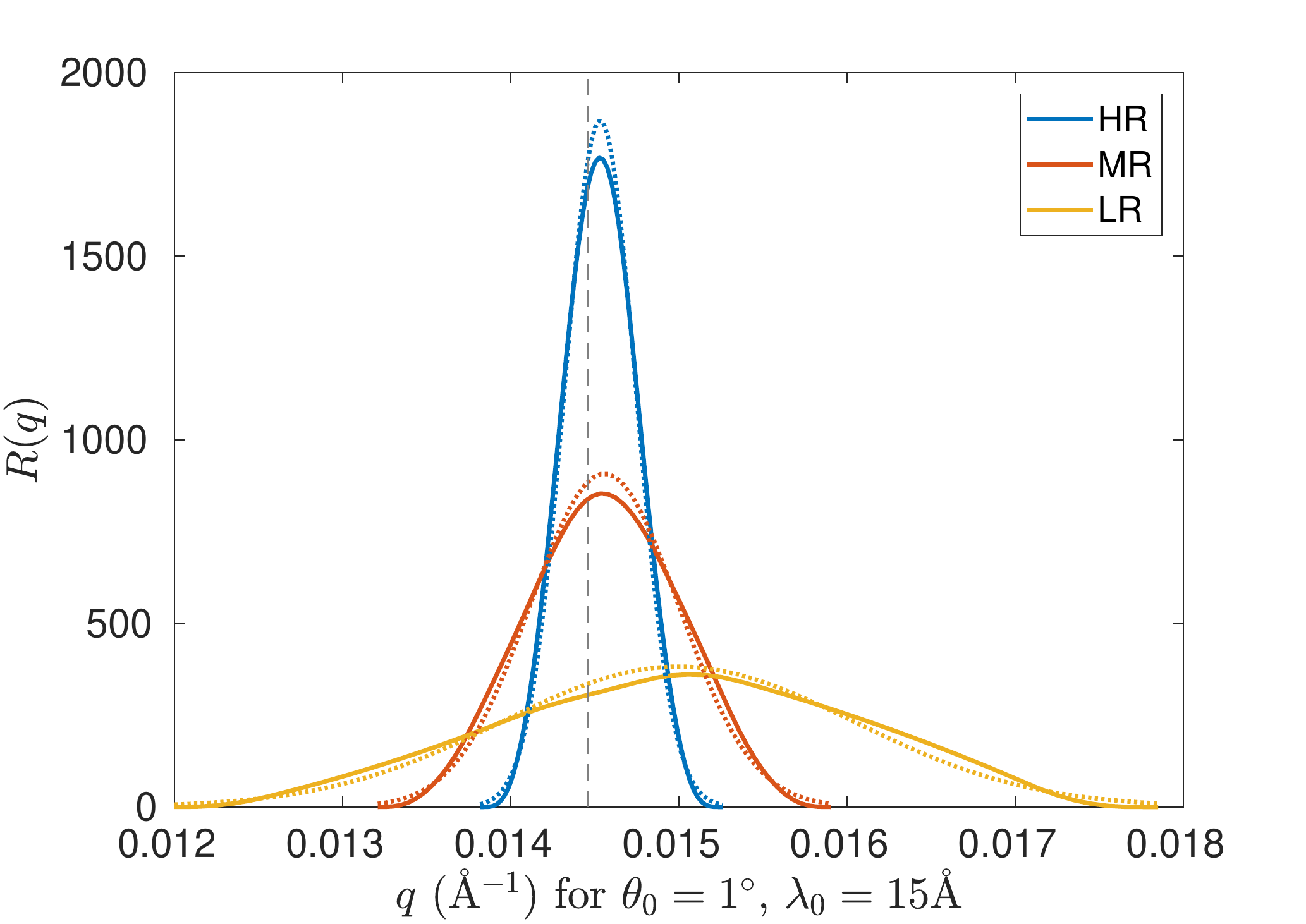}
\caption{Exact resolution $R(q)$ (full line) vs.  transfer vector $q$ for $\theta_0=1^\circ$ and $\lambda_0=15$\AA~ (crosssection of 3D plot of Fig.\ref{figR}) for the three configurations of Table\,\ref{tab1}. Dotted lines are gaussian curves of same mean and standard deviation.}
\label{figcompaR}
\end{figure}

For the sake of simplicity, let us first deal with the ``small angle approximation'' $\sin (\theta)\simeq \theta$. Then, using Eq.\ref{eqtheta} the transfer vector rewrites\,:
\begin{equation}\label{defqbis}
q=\frac{4\pi \theta}{\lambda} =4\pi\frac{\theta_0+\alpha-\gamma}{\lambda}
\end{equation}
where $\lambda$ and $\alpha$ are random variables with densities $H(\lambda)$ and $P(\alpha)$, respectively, whereas $\gamma=-(c\lambda)^2$.
Let us denote $M(q)$ the measurement of the physical quantity $m(q)$. Quite generally, for one time-of-flight channel at $q_0=4\pi\theta_0/\lambda_0$ one can write
\begin{equation}\label{mconvolution}
M(q_0)=\int\diff\lambda H(\lambda)\int\diff \alpha P(\alpha) \quad m\left({4\pi\frac{\theta_0+\alpha-\gamma}{\lambda}}\right)
\end{equation}
Using $\alpha=(\lambda q/4\pi)-\theta_0-(c\lambda)^2$ and $\diff\alpha=(\lambda/4\pi) \diff q$, one gets: 
\begin{equation}\label{mconvolution2}
\begin{array}{rl}
M(q_0)&=\displaystyle\int \diff\lambda H(\lambda) \int\diff q \frac{\lambda}{4\pi}  P\left({\frac{\lambda q}{4\pi}-\theta_0 -(c\lambda)^2}\right)  m(q)\\
&=\displaystyle\int \diff q m(q) \int \diff\lambda H(\lambda) \frac{\lambda}{4\pi} P\left({\frac{\lambda q}{4\pi}-\theta_0 -(c\lambda)^2}\right)
\end{array}
\end{equation}
By definition, the last integral is the resolution function of the transfer vector $q$\,:
\begin{equation}\label{eqR}
R(q) = \int \diff\lambda H(\lambda) \frac{\lambda}{4\pi} P\left({\frac{\lambda q}{4\pi}-\theta_0 -(c\lambda)^2}\right)
\end{equation}
The same procedure can be used for the exact expression $q={4\pi\sin(\theta_0+\alpha-\gamma)}/{\lambda}$ leading to $\alpha=\sin^{-1}(\lambda q/4\pi)-\theta_0-(c\lambda)^2$ and $\diff\alpha =\lambda \diff q/\sqrt{(4\pi)^2-(\lambda q)^2}$. One obtains\,:
\begin{equation}\label{eqR2}
R(q) = \int \diff\lambda \frac{H(\lambda) \lambda  P(\sin^{-1}(\lambda q/4\pi)-\theta_0-(c\lambda)^2)}{4\pi\sqrt{1-(\lambda q/4\pi)^2}}
\end{equation}
Eq.\ref{eqR2} can be numerically calculated as is (see python code \url{https://bitbucket.org/LLBhermes/pytof/}) from Eq.\ref{D}, \ref{eqH} and \ref{eqP}.
In Fig.\ref{figR}, the whole resolution $R(q)$ is plotted for the three typical configurations as a function of the time-of-flight channel for $\theta_0=1^\circ$. Firstly, notice the shift of the mean transfer vector value $\bar q=\int q R(q) \diff q$ compared to $q_0$. This shift results from the wavelength distribution of the incident beam (Fig.\ref{figl}) and from gravity (Fig.\ref{figgbar}). Both effects contribute to the same result for $\theta_0>0$ but oppose if $\theta_0<0$. To properly account for this effect, the simplest and more accurate way is to compute the exact resolution function.
Secondly, notice that the profile of the exact resolution function differs from its gaussian approximation (see Fig.\ref{figcompaR}). In any case, the exact resolution function has a compact support (\textit{i.e.} $R$ is non-zero inside a closed and bounded set of $q$-values), unlike gaussian curve that should be numerically cut beyond an arbitrary number of times the standard deviation. For the configuration allowing a wider resolution, the differences between exact resolution and gaussian approximation are much more important and even the modal-values differ. All these differences cannot reasonably be taken into account summarizing the resolution with mean and standard deviation values.

\newpage
\section{Application}

Let us consider the measurement $M_k$ (performed with statistical errors bars $\sigma_k$ with $k\in\{1, 2,\dots N\}$) of the reflectivity of a sample over $N$ time-of-flight channels. We denote $m(q,\textbf p)$ the theoretical reflectivity that depends on $n$ physical parameters that make the coordinates of the vector $\textbf p$. For a given time-of-flight channel $k$ of resolution function $R_k$, the expected theoretical measurement $M_{\textrm{th,}k}$ is given by\,:
\begin{equation}\label{eqexpectation}
M_{\textrm{th,}k}(\textbf p)=\displaystyle\int\! m(q,\textbf p) R_k(q) \diff q 
\end{equation}
Let us define the mean relative distance $\chi^2$ per channel between $M$ and $M\subscr{th}$ as
\begin{equation}\label{eqchi2}
\chi^2(\textbf p)=\frac{1}{N}\sum_{k=1}^N\left({
\frac{M_k-M_{\textrm{th,}k}(\textbf p)}
{\sigma_k}
}\right)^2
\end{equation}
The parameter-vector $\textbf p$ can be experimentally determined by minimizing $\chi^2$ following a standard numerical optimization procedure (curve fitting). The correctness of the resolution function can thus be evaluated\,: 1)~from the correctness of the so determined parameters values; and 2)~from the correctness of the final matching between $M$ and $M\subscr{th}$ (low value for~$\chi^2$ and no correlation in the residual $(M-M\subscr{th})/\sigma$).

\begin{figure}
\begin{center}
\includegraphics[width=\wb\linewidth]{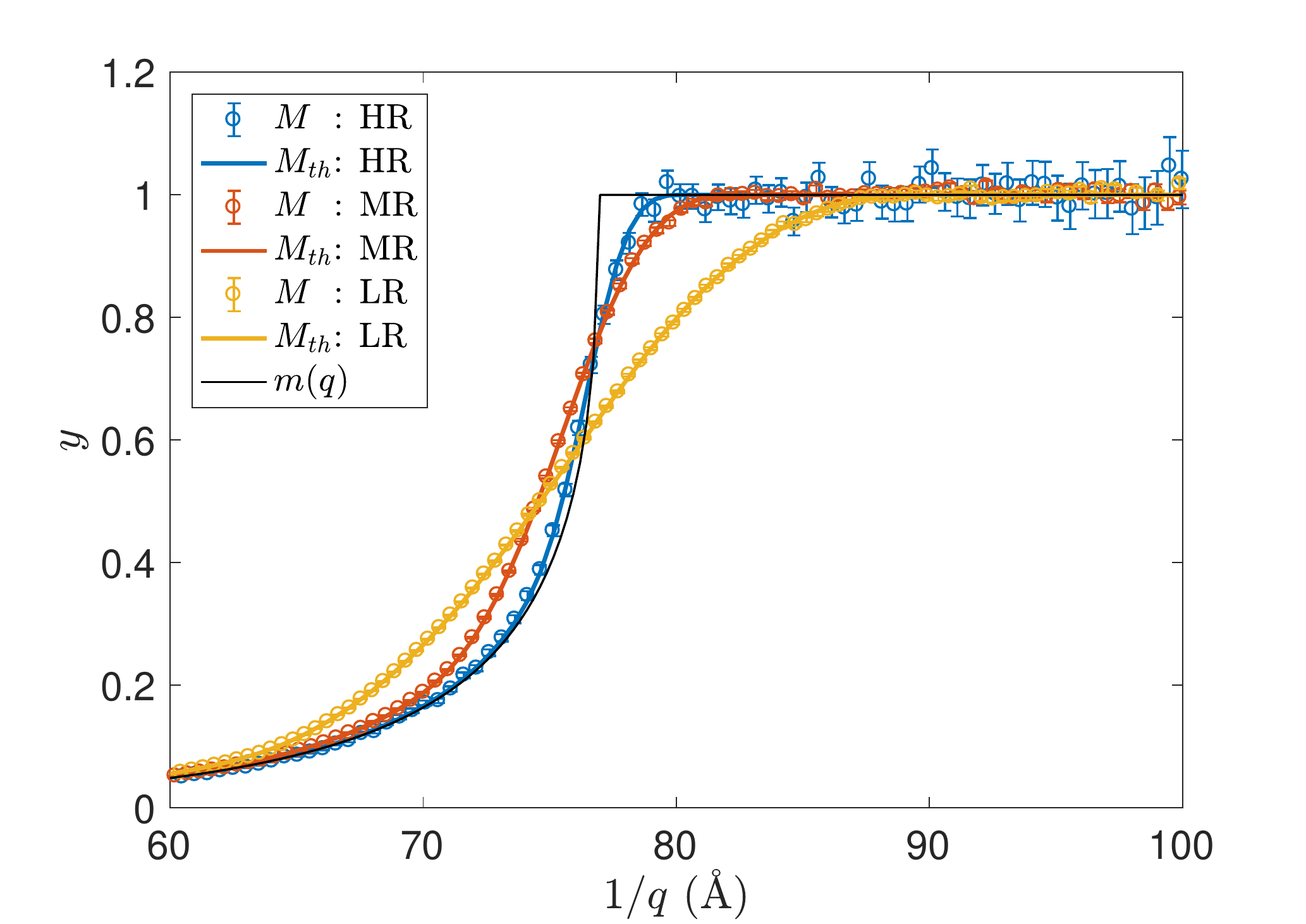}\\
\includegraphics[width=\wb\linewidth]{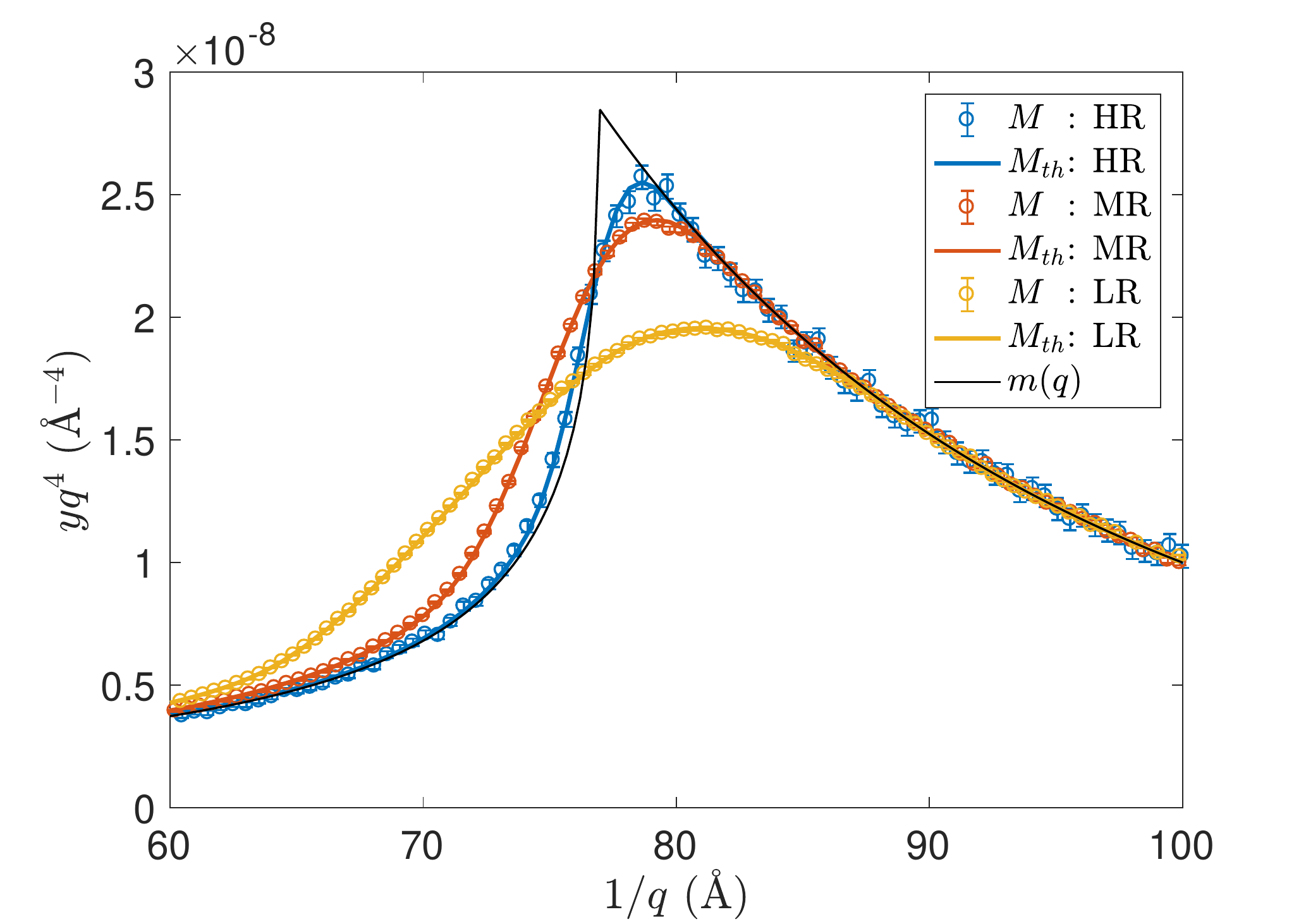}
\caption{Reflectivity for an amorphous silica block in the region of the edge of the total reflection. Top\,:~measurement $M$, best fit $M\subscr{th}$ (Eq.\,\ref{eqexpectation}) and model $m(q)$ (Eq.\,\ref{eqFresnel}) vs. $1/q$ (because channels are almost regularly spaced in wavelength). Bottom: same data with $y$-coordinates multiplied by $q^4$.
For these best fits $q_c=1.298, 1.306, 1.292\times10^{-2}$\AA$^{-1}$ and $\chi^2=1.11$, 3.97 and 3.51 for the high-resolution (HR), medium-resolution (MR) and low-resolution (LR), respectively.}
\label{figex1a}
\end{center}
\end{figure}

For such an evaluation, sample-candidate should display a strong variation of reflectivity in the accessible $q$-range, in order to maximize the effect of convolution by the resolution function (Eq.\,\ref{eqexpectation}), with a minimum number of unknown parameters (coordinates of $\bf{p}$).
From this point of view, the reflectivity near the total reflection plateau of the interface between air and a smooth and pur solid is likely the most adequate. We have choosen an amorphous silica block with a polished surface, which reflectivity writes\,:
\begin{equation}\label{eqFresnel}
m(q)=\left({\frac{q-\sqrt{q^2-q_c^2}}{q+\sqrt{q^2-q_c^2}}}\right)^2\times e^{-\sigma_h^2 q\sqrt{q^2-q_c^2}}
\end{equation}
The first term of this product is the Fresnel's reflectivity of a perfectly flat surface, where $q_c=(16\pi\rho)^{1/2}$ is the edge of the total reflectivity plateau and $\rho$ the scattering length density of amorphous silica. Whereas the second term accounts for the surface roughness of characteristic height $\sigma_h$. Measurements were done in the three configurations of Table\,\ref{tab1} for $\theta_0=1^\circ$. Results and data fitting are plotted in Fig.\ref{figex1a}.
For the three configurations, the values for $\chi^2$ at the optimum are very small and the values determined for $q_c$ are in very good agreement and consistent with the density of amorphous silica (here $\bar q_c=(1.300\pm0.006)\times 10^{-2}$\AA$^{-1}$ leads to $(2.13\pm0.02)$\,g/cm$^3$ for the density of amorphous silica). Note that the increase of $\chi^2$ with the broadening of resolution is simply due to the gain of flux resulting in smaller statistical error bars $\sigma_k$ in Eq.\ref{eqchi2}.

\begin{figure}
\begin{center}
\includegraphics[width=\wb\linewidth]{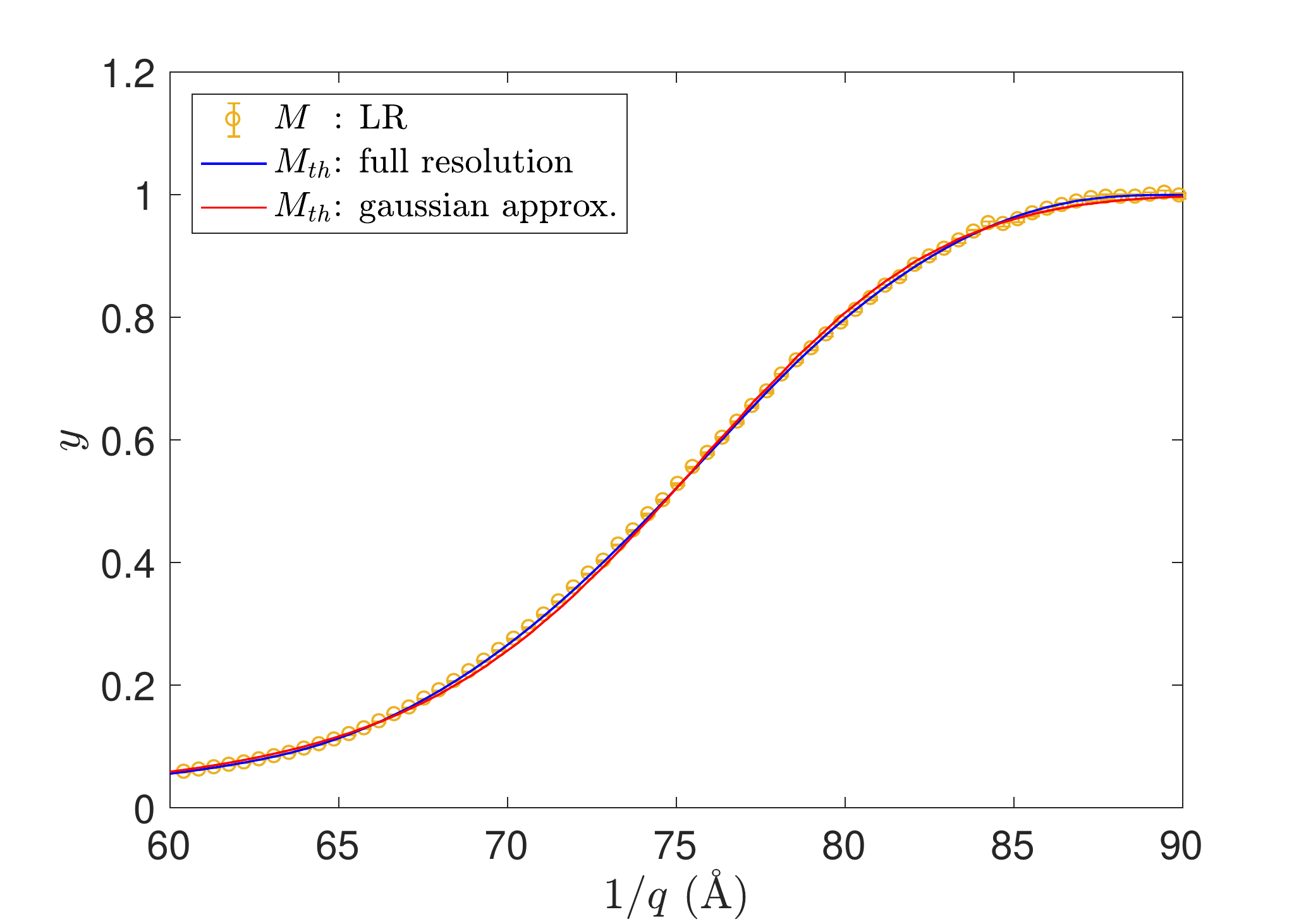}\\
\includegraphics[width=\wb\linewidth]{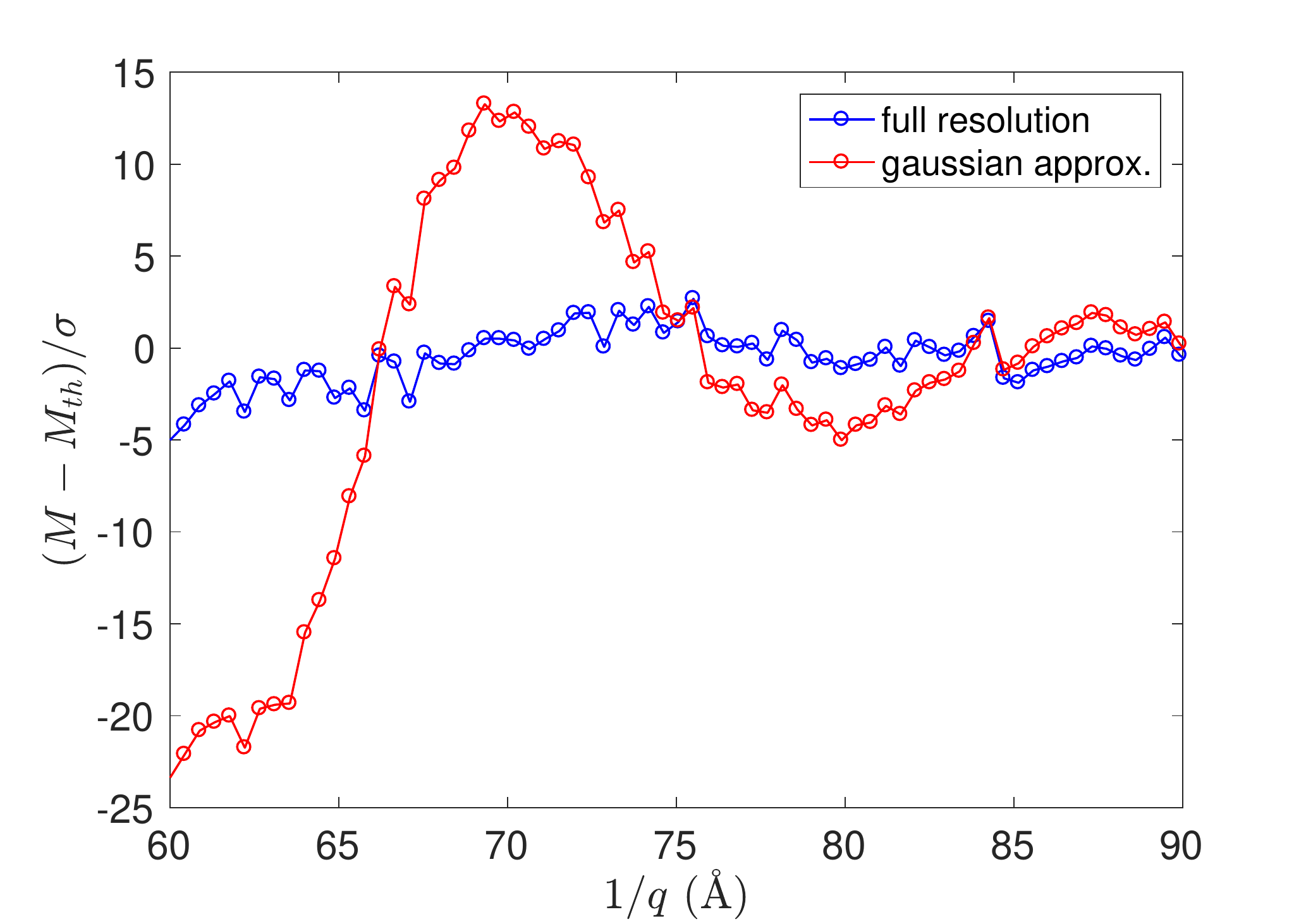}
\caption{Reflectivity for an amorphous silica block in the region of the edge of the total reflection plateau. Top\,:~measurement $M$ and best fits $M\subscr{th}$ using the exact resolution or its gaussian approximation (computed over an interval half-width equal to 3 times the standard deviation). Bottom\,:~$(M-M_{th})/\sigma$. The use of the exact resolution leads to $\chi^2=3.51$, its gaussian approximation to $\chi^2=218$.}
\label{figex2}
\end{center}
\end{figure}

The use of the exact resolution function as plotted in Fig.\ref{figR}, should appear unnecessarily tricky compared to the use of a gaussian curve of same average and standard deviation.
This is not the case for two reasons: 1)~the calculation of the exact resolution function is anyway needed to determine the average and standard deviation values of its  gaussian approximation curve; 2)~once this is achieved, convolution by the exact resolution or its gaussian approximation requires the same computing power.
Also, convolution by the exact resolution function will always produces better results in particular at low resolution.  Best fits obtained using the gaussian approximation (computed over a support-half-width equal to 3 times the standard deviation) of the exact resolution function were done for the examples given in Fig.\ref{figex1a}. The corresponding $\chi^2$-values are found to be 1.54, 15.1 and 218 (instead of  $\chi^2=1.11$, 3.97 and 3.51 with the exact function) for the high-resolution (HR), medium-resolution (MR) and low-resolution (LR), respectively.
In Fig.\ref{figex2}-top the results for the two fits are compared for the low resolution configuration of Table\,\ref{tab1}. Differences are subtle but are emphasized by plotting the residual $(M-M_{th})/\sigma$ (Fig.\ref{figex2}-bottom) that clearly shows correlations in the case of the gaussian approximation. It is important to emphasize that the gaussian approximation we are discussing here should not be confused with a resolution function that would be calculated following the central-limit theorem \textit{i.e.} by adding the variances of each term. Here, as the mean and standard deviation values are correctly calculated (from the exact resolution function), fitting using one or the other resolution-curve does not lead to significant difference in parameters best-values, nor in their confidence intervals, although the $\chi^2$-values are strongly different. However, the gaussian resolution-curve that could be produced by applying the central limit theorem leads to much less satisfying parameters values.
The computation of the exact resolution profile is thus mandatory to reach the most accurate data fitting results.

By definition, using the exact resolution function is more accurate than using its gaussian approximation, but it also saves time. Computation of the exact resolution function is done only once during data reduction. The  time spent is negligible compared to the time needed for data fitting that consists in many iterations of convolution of a theoretical model by the resolution. The exact resolution function has a compact support whereas its gaussian approximation has not. Thus for a given high percentile, the gaussian approximation needs a more extended sampling and thus is more time consuming during the data fitting stage than the exact function. This removes a lot of interest in the approximation.

\section{Conclusion}

In this paper we present the calculation of the exact and comprehensive resolution function for a time-of-flight neutron reflectometer in a way that accounts for all contributions without any assumption of the gaussian distribution or independence  of the corresponding variables. The step-by-step procedure comes with a fully documented Python module (\url{https://bitbucket.org/LLBhermes/pytof/}), whose routines match with the equations and terminology of the paper. This module that can be easily used for numerical applications to any specific case, from the computation of the resolution to data-fitting. In addition, this module allows the reader to reproduce most of the figures of this paper and to change easilly their parameters.

We have shown that in case the resolution is relaxed, the resulting resolution function departs strongly from a gaussian profile and that using the exact function provides much more accurate results. This point will be highly relevant with the emergence of compact and low-flux neutron sources (see \textit{e.g.}\,\cite{ott:cea-01873010}) which will likely require such relaxed resolutions.

Here we tried to treat the different contributions on the resolution of  time-of-flight neutron refolectometer in an exhaustive but still general manner. However, some peculiar points, which would require some special attention, have not been examined, mainly\,: detector resolution and pixel-binning\,\cite{Cubitt:ge5018} in case a position sensitive detector is used; non-flat samples\,\cite{Cubitt:ge5018}... We hope this paper will help to manage these particular cases.

\bibliographystyle{plain}

\end{document}